\documentclass[prd,amsmath,notitlepage,eqsecnum,twocolumn]{revtex4-2}

\pdfoutput=1
\usepackage[utf8]{inputenc}

\usepackage{graphicx}
\usepackage{float}
\usepackage{epstopdf,cancel}
\usepackage{epsf,latexsym,bbm,euscript}
\usepackage{amssymb,amsmath}
\usepackage{mathtools}
\usepackage{times,graphics}
\usepackage{soul,xcolor}
\usepackage{mathtools}
\usepackage[normalem]{ulem}
\usepackage{setspace}
\usepackage{enumitem}
\usepackage{accents}
\usepackage{booktabs}
\usepackage{microtype}
\usepackage[caption=false]{subfig}
\usepackage{etoolbox}
\usepackage{titlesec}
\usepackage{siunitx}

\makeatletter
\newcommand*{\defeq}{\mathrel{\rlap{%
			\raisebox{0.3ex}{$\m@th\cdot$}}%
		\raisebox{-0.3ex}{$\m@th\cdot$}}%
	=}
\newcommand*{\eqdef}{=\mathrel{\rlap{%
			\raisebox{0.3ex}{$\m@th\cdot$}}%
		\raisebox{-0.3ex}{$\m@th\cdot$}}%
}
\makeatother

\makeatletter
\g@addto@macro\bfseries{\boldmath}
\makeatother

\makeatletter
\def\thesubsection{\thesection.\arabic{subsection}}
\def\p@subsection{}
\makeatother

\titleformat{\section}[hang]
{\normalfont\normalsize\bfseries\MakeUppercase}
{\thesection.}{0.75em}
{\raggedright}

\titleformat{\subsection}[hang]
{\normalfont\normalsize\bfseries}
{\hspace{1em}\thesubsection.}{0.5em}
{\hspace{0pt}\raggedright}

\titleformat{\subsubsection}[hang]
{\normalfont\normalsize\bfseries}
{\thesubsubsection.}{1em}{\raggedright}

\titlespacing*{\section}{0pt}{4mm}{1mm}
\titlespacing*{\subsection}{0pt}{3mm}{0.5mm}
\titlespacing*{\subsubsection}{0pt}{2.25mm}{0.25mm}

\apptocmd{\appendix}{%
	\titleformat{\section}[hang]
	{\normalfont\normalsize\bfseries}
	{Appendix \thesection:}{0.5em}{\raggedright}%
	\titlespacing*{\section}{0pt}{3mm}{0.5mm}%
}{}{}

\usepackage{scalerel}
\usepackage{tikz}
\usetikzlibrary{svg.path}
\definecolor{orcidlogocol}{HTML}{A6CE39}
\tikzset{
	orcidlogo/.pic={
		\fill[orcidlogocol] svg{M256,128c0,70.7-57.3,128-128,128C57.3,256,0,198.7,0,128C0,57.3,57.3,0,128,0C198.7,0,256,57.3,256,128z};
		\fill[white] svg{M86.3,186.2H70.9V79.1h15.4v48.4V186.2z}
		svg{M108.9,79.1h41.6c39.6,0,57,28.3,57,53.6c0,27.5-21.5,53.6-56.8,53.6h-41.8V79.1z M124.3,172.4h24.5c34.9,0,42.9-26.5,42.9-39.7c0-21.5-13.7-39.7-43.7-39.7h-23.7V172.4z}
		svg{M88.7,56.8c0,5.5-4.5,10.1-10.1,10.1c-5.6,0-10.1-4.6-10.1-10.1c0-5.6,4.5-10.1,10.1-10.1C84.2,46.7,88.7,51.3,88.7,56.8z};
	}
}
\newcommand\orcidlink[1]{\href{https://orcid.org/#1}{\mbox{\scalerel*{
				\begin{tikzpicture}[yscale=-1,transform shape]
				\pic{orcidlogo};
				\end{tikzpicture}}{X}}}}

\def\sg{\textsl{g}}
\def\cO{\mathcal{O}}

\newcommand{\be}{\begin{equation}}
\newcommand{\ee}{\end{equation}}
\newcommand{\ba}{\begin{eqnarray}}
\newcommand{\ea}{\end{eqnarray}}
\def\half{\tfrac{1}{2}}

\def\6{{\langle}}
\def\9{{\rangle}}
\newcommand{\pad}{\partial}

\def\rin{\text{in}}
\def\rou{\text{out}}

\def\gb{\mathfrak{b}}
\def\gc{\mathfrak{c}}
\def\gd{\mathfrak{d}}

\def\ab{\mathsf{b}}
\def\al{\mathsf{l}}
\def\an{\mathsf{n}}
\def\am{\mathsf{m}}
\def\ave{\mathsf{e}}
\def\aw{\mathsf{w}}
\def\au{\mathsf{u}}

\def\mfa{\mathfrak{a}}

\def\aA{\mathsf{A}}

\newcommand\eA{{\EuScript{A}}}
\newcommand\eC{{\EuScript{C}}}
\newcommand\eB{{\EuScript{B}}}
\def\eF{\EuScript{F}}

\def\bfe{\mathbf{e}}
\def\bfl{\mathbf{l}}

\usepackage{fancyhdr}
\usepackage{url,hyperref}
\hypersetup{colorlinks,linkcolor={blue!55!black},citecolor={red!50!black},urlcolor={blue!45!black},breaklinks=true}

\begin{document}
	
\title{Gravity-induced birefringence in spherically symmetric spacetimes}
		
\author{Sebastian Murk\orcidlink{0000-0001-7296-0420}}
\email{sebastian.murk@oist.jp}
\affiliation{Quantum Gravity Unit, Okinawa Institute of Science and Technology, 1919-1 Tancha, Onna-son, Okinawa 904-0495, Japan}

\author{Daniel R.\ Terno\orcidlink{0000-0002-0779-0100}}
\email{daniel.terno@mq.edu.au}
\affiliation{School of Mathematical and Physical Sciences, Macquarie University, New South Wales 2109, Australia}

\author{Rama Vadapalli\orcidlink{0000-0001-7470-1342}}
\email{venkataramaraju.vadapalli@students.mq.edu.au}
\affiliation{School of Mathematical and Physical Sciences, Macquarie University, New South Wales 2109, Australia}
	
\begin{abstract}
	Geometric optics effectively describes the propagation of electromagnetic waves when the wavelength is much smaller than the characteristic length scale of the medium, making wave phenomena like diffraction negligible. As a result, light propagation in a vacuum is typically modeled by rays that follow null geodesics. However, general relativity predicts that polarization-dependent deviations from these geodesics occur in an inhomogeneous gravitational field. In this article, we evaluate the corrections for the deflection and emission of light by a massive gravitating body. Additionally, we derive the scaling behavior of the physical parameters characterizing the trajectories. The calculations are performed at leading order in frequency. We use these results to assess the significance of the birefringence effect in various astrophysical observations. We find that the effect cannot be measured with current instruments but may be detectable in the near future.
\end{abstract}
	
\maketitle
\thispagestyle{fancy}
	
\section{Introduction}\label{s:intro}
The study of optical phenomena ranks among the oldest scientific pursuits, predating the emergence of the electromagnetic theory by many centuries. Its formulation completed the classical view of our universe and laid the foundations for quantum mechanics and relativity. Yet, in modern classical and quantum optics \cite{BW:99,MW:99},  wave propagation is often analyzed using ray tracing, with corrections applied when necessary. Although general relativity remains a classical theory, it reshapes our flat-spacetime view of wave propagation: the gravitational fields of massive bodies bend light rays, rotate their polarization, and render the vacuum birefringent.

As long as the electromagnetic field intensity is low enough for nonlinear effects of quantum electrodynamics and the backreaction on spacetime geometry via the Einstein field equations to be negligible, the propagation is governed by the classical wave equations on a fixed curved background~\cite{MTW,H:19}.
 For the minimally coupled electromagnetic field, the vector potential $A^\mu$ satisfies
\begin{equation}
	\Box A^\mu- R^\mu_{\ \nu} A^\nu=0 , \label{wave1}
\end{equation}
where the d'Alembert operator is defined via $\Box \! \defeq \! \nabla^\mu \nabla_\mu$, $\nabla_\mu$ and $R^\mu_{\ \nu}$ denote the covariant derivative and Ricci tensor associated with the background metric $g_{\mu\nu}$, respectively, and we impose the Lorenz gauge $\nabla_\mu A^\mu = 0$. The order-by-order solutions are derived by considering a decomposition of the vector potential as
\begin{equation}
	A^{\mu}(x) =\eA^\mu(x)e^{i\Phi(x)}, \quad \eA^\mu(x)=\sum_{n= 0}^{{\infty}} {{\omega}}^{-n} A^\mu_n(x) , 	\label{asy_exp}
\end{equation}
where $\Phi$ is the phase (or eikonal function), the amplitudes {$\eA^\mu$} are slowly varying on the relevant timescales, and the large parameter ${{\omega}}$ is  related to the peak frequency  of the solution~\cite{MTW,BW:99,H:19}. In our convention the eikonal function implicitly depends on it via
\begin{equation}
	k_\mu \defeq  \nabla_\mu \Phi \equiv  \pad_\mu\Phi\eqdef \omega l_\mu ,
	\label{wav}
\end{equation}
where $k_\mu$ denotes the wave vector. The eikonal and the amplitudes can be determined from the equations for the coefficients of {the} various ${{\omega}}^{-n}$ terms  that are obtained by inserting this vector potential into the wave equation.

Substitution of the lowest-order term $\cO(\omega^0)$ of the decomposition \eqref{asy_exp} into Eq.~\eqref{wave1} results in the propagation equations for the wave vector and its polarization. These provide the basis for the formulation of geometric optics and the gravitational Faraday effect (also known as the Rytov-Skroski\u{\i} effect).

The wave vector \cite{BW:99,MTW,H:19} $k_\mu=\omega l_\mu$ defines the propagation and spatial periodicity of the wave. In the standard approach to the eikonal equation it is not expanded in inverse powers of $\omega$. Then, it is null in all orders of the asymptotic expansion of Eq.~\eqref{asy_exp}, and thus satisfies the eikonal equation
\begin{equation}
	\al^2 = \pad_\mu \Phi \pad^\mu \Phi/\omega^2=0 ,
	\label{eq_eik}
\end{equation}
which is a restatement of the null condition in terms of the phase function. It is the Hamilton-Jacobi equation for massless particles on a given background spacetime. As such, it is equivalent to a dynamical system of massless point particles described by a Hamiltonian $H(l_\mu,x^\mu)$ \cite{BW:99,A:89}. These fictitious particles are often referred to as photons, even if the context is purely classical. In asymptotically flat spacetimes, we resolve the ambiguity in the definition of $\al$ by requiring that $\omega$ is the frequency observed at infinity, i.e., $l^t\to 1$ in the asymptotically flat region.

The three-dimensional hypersurfaces of constant $\Phi$ are null. The hypersurface-orthogonal integral curves of $l^\mu$ form a twist-free null geodesic congruence. These geodesics are the light rays of geometric optics. Alternatively, they may be interpreted as the trajectories of fictitious classical photons that generate the hypersurface of constant phase $\Phi$ and at the same time are orthogonal to it due to Eq.~\eqref{eq_eik}~\cite{MTW}.

The spacelike polarization vector $\ave$ is transversal to the null geodesic generated by $\al$. Both vectors are parallel-propagated according to \cite{C:92,MTW}
\begin{equation}
	\nabla_\al\al=0 , \qquad \nabla_\al\ave=0 .
	\label{paral}
\end{equation}
These equations are obtained from the two leading terms resulting from the substitution of the expansion \eqref{asy_exp} into the propagation equation \eqref{wave1}. The first equation is responsible for the deflection of light, and the second for the polarization rotation.

The deflection of light from the fiducial Euclidean path is the first classical test of general relativity \cite{MTW,W:18}. Gravitational Faraday rotation was found to dramatically alter the polarization of X-ray radiation emitted from the accretion disk of the black hole candidate in Cyg X-1 \cite{SC:77}. Notably, accounting for this effect played a crucial role in the polarization analysis of the emission spectrum of the black hole candidate situated in M87 \cite{EHT:23}. It has also been investigated in the context of gravitational lensing \cite{KDBJ:91,BDKR:04,KAM:24} and interactions of gravitational and electromagnetic waves \cite{F:08,D:21}. The interpretation of these results involves subtleties that are related to both the differences in superfluously similar physical situations and the important role that is played by the choice of reference frames \cite{FL:82,F:08,BT:11,BDT:11,S:21}.

If the wave vector is expanded to the first order in $\omega^{-1}$, then the second of the leading terms of the propagation equations also includes the  term in the expansion of Eq.~\eqref{asy_exp} of the same order.\ In App.~\ref{app:shortwave}, Eqs.~\eqref{app:eq:prop2}--\eqref{app:eq:Amu}, we show that this systematic procedure leads to the effective dispersion relations and the corrected phase that are used in Refs.~\cite{F:20,O:20}.\ This term causes the propagation of left- and right-handed circularly polarized components of a beam of light along different paths in an inhomogeneous gravitational field.\ This phenomenon is referred to as gravitational birefringence or gravitational spin Hall effect \cite{S:21,M:74,FS:11,DS:17,F:20,O:20,AO:23,S:24,FS:12}. Several approaches can be used to evaluate this effect. Efficient schemes that are applicable for general spacetimes have been developed recently and allow us to establish relations between different approaches \cite{F:20,O:20,HO:22}.

Even in the simplest spacetimes (e.g., Schwarzschild or extremal Kerr black holes, or Robertson-Walker spacetimes), numerical simulations reveal some interesting consequences of the ostensibly small deviations from geometric optics \cite{FS:12,DS:17,O:20}. The perturbative scale is set by the parameter $1/(\omega L)$, where $L$ is a typical length scale (e.g., the Schwarzschild radius $r_\sg \equiv 2M$).\ Several analytical investigations \cite{GBM:07,DMS:19,DT:23} have identified some of these features.

Using the formulation developed by Frolov in Ref.~\cite{F:20}, we investigate the effects of gravity-induced birefringence in the Schwarzschild spacetime and perform calculations at the leading order of $1/(\omega r_\sg)$. For impact parameters larger than the critical value $3\sqrt{3} r_\sg/2$, we obtain explicit expressions for the quantities that characterize polarization-dependent orbits. At the same order $1/\omega$, they scale as various powers of the dimensionless ratio $\ell/r_\sg$, where $\ell \defeq j/\varepsilon$ denotes the impact parameter, and $j$ and $\varepsilon$ the conserved angular momentum and the conserved energy of the photons, respectively.

The remainder of this article is organized as follows: We present the basic equations underlying our approach in Sec.~\ref{equation}, including their numerical solutions and the schematics of obtaining iterative analytical solutions. Results pertaining to the scaling of various orbital parameters are presented in Sec.~\ref{solutions}. In Sec.~\ref{astro}, we outline applications of our findings for astrophysical observations. Lastly, in Sec.~\ref{sec:discussion}, we discuss the physical implications of our results and survey prospects for future directions in this research domain. Additional mathematical details are provided in the appendices, and the \textsc{Mathematica} code detailing explicit calculations is openly available in the \textsc{GitHub} repository listed as Ref.~\cite{GitHubRep}.

We use the $(-,+,+,+)$ metric signature and geometrized units where $c=G=1$. Index-free four-vectors are denoted by the Sans Serif font, e.g., $\al$, $\ave$, $\aw$, and by the Computer Modern font (the default font family in \LaTeX) otherwise, e.g., $l^\mu$, $e^\mu$, $w^\mu$. {The labels for vectors in an orthonormal tetrad are enclosed by parentheses whenever any of their co- and/or contravariant components are referred to explicitly [cf.\ Eq.~\eqref{eq:master}]. If no components are referenced explicitly we omit the parentheses to reduce notational clutter [cf.\ Eqs.~\eqref{eq:nt-d} and \eqref{eq:m.mbar}].}
Three-vectors are indicated by boldface, e.g., $\mathbf{l}$ and $\bfe$. Greek indices are assumed to run from $0$ to $3$.

\section{Basic equations} \label{equation}
Our starting point is the system of propagation equations \cite{F:20} for the tangent to a null ray  $l^\mu\defeq dx^\mu/d\tau$ (where $\tau$ denotes the affine parameter) and three additional vectors that together form a null tetrad \cite{C:92}. Unlike Ref.~\cite{F:20}, we use the two real linear polarization vectors instead of the complex circular polarization vectors.  This choice simplifies the analysis and reduces errors of numerical calculations.

The polarization four-vectors satisfy
\begin{align}
	 \ave_i^2=1, \quad \al\cdot\ave_i=0, \quad \an\cdot\ave_i=0,	
	 \label{eq:nt-d}	
\end{align}
where $i=1,2$ and the auxiliary null vector $\an$ satisfies $\al\cdot\an=-1$. The Newman-Penrose \cite{NP:62,C:92} null tetrad $(\al,\an,\am,\bar \am)$ is formed by setting
\begin{align}
	\am=\frac{1}{\sqrt{2}}(\ave_1+i\ave_2), \quad \bar \am=\frac{1}{\sqrt{2}}(\ave_1-i\ave_2).
	\label{eq:m.mbar}
\end{align}
A polarized light ray follows a null but in general nongeodesic trajectory whose acceleration in the high-frequency limit is given by
\begin{align}
	\frac{D^2x^\mu}{D\tau^2}=-\frac{\sigma}{\omega}R^\mu_{~\nu\alpha\beta} l^\nu e_{(1)F}^\alpha e_{(2)F}^\beta\eqdef w^\mu . 	
	\label{eq:master}
\end{align}
Here $R^\mu_{~\nu\alpha\beta}$ denotes the Riemann tensor and $\sigma=\pm 1$ corresponds to right/left circular polarization. The derivation of this expression requires that the tetrad is propagated according to
\begin{align}
	\nabla_\al\an_F=0, \quad \nabla_\al \am_F=-\kappa \an_F, \quad \nabla_\al \bar\am_F=-\kappa^* \an_F,
	\label{tetrad-prop}
\end{align}
where the acceleration parameter $\kappa$ is given by
\begin{align}
	\kappa \defeq -\aw\cdot\am =-\frac{i\sigma}{\omega}R_{\mu\nu\alpha\beta} l^\mu m_{F}^\nu m_{F}^\alpha \bar m_{F}^\beta. \label{F-prop}
\end{align}
This set of equations guarantees that the relations between the tetrad vectors are preserved along the trajectory $x^\mu(\tau)$, and the subscript $F$ is used to distinguish such tetrads (and to indicate that they satisfy the analog of Fermi-propagation that is adapted to null trajectories). Using the volume form in terms of the tetrad vectors (see App.~\ref{app:null.tetrads}), we find that
\begin{align}
	{e_{(1)[\mu} e_{(2)\nu]}} = -\sqrt{-g}\epsilon_{\mu\nu\rho\sigma}l^\rho n^\sigma,
\end{align}
and thus it is sufficient to parallel propagate $\an$ along the trajectory to obtain the acceleration $\aw$ of Eq.~\eqref{eq:master}.

The conditions \eqref{eq:nt-d} and \eqref{F-prop} uniquely specify the tetrad up to three distinct types of transformations (see App.~\ref{app:null.tetrads}) \cite{C:92} at the initial point of the trajectory. This prompts the question of how uniquely deviations from the geodesic trajectory driven by the acceleration of Eq.~\eqref{eq:master} are defined, and how possible ambiguities are to be interpreted \cite{F:20,O:20,HO:22}. If the meaning of $\al$ as a tangent is to be maintained, then only a rescaling of the affine parameter is allowed, which has no bearing on the modified trajectory. We set the scale in Eq.~\eqref{eq:lt} below. We also note that the ambiguity in the definition of the individual polarization vectors within the polarization plane [transformations of type IIIb, see App.~\ref{app:null.tetrads}, Eq.~\eqref{app:eq:A.IIIb}] does not affect the propagation equations.

In the plane wave idealization, the polarization plane --- whether for the entire beam or its individual Fourier components --- has an operational meaning as the plane defined by the two three-vectors of the electric field that correspond to right- and left-circular polarization. Mathematically, it is introduced as follows \cite{PT:04}: Using an orthogonal tetrad (whose timelike vector is naturally associated with the four-velocity of a particular observer \cite{MTW,BT:11,O:20,HO:22}), we write $\al=\big(l^{(0)},\bfl\big)$ and the electric field $\mathbf{E}$ satisfies the transversality condition $\bfl\cdot \mathbf{E}=0$. The two directions that correspond to linear polarizations (set, for instance, by adopting the Wigner construction of the induced representation of the Poincar\'{e} group \cite{PT:04,BDT:11}) are the three-vectors $\bfe_{1,2}$ that are promoted to the four-vectors $\ave_{1,2}=(0,\bfe_{1,2})$. Under Lorentz transformations (and, in particular, boosts) the three-dimensional transversality relation is preserved, and the polarization plane has a well-defined transformation law. In fact, transformations of $\al$ and $\am$ form a realization of the Wigner little group \cite{LPT:03}. However, since  {$e_{(1)[\mu}e_{(2)\nu]}$} does not transform in the explicitly Lorentz-covariant fashion, accelerations and thus deviations from the geodesic trajectory according to different observers are different \cite{HO:22}. In what follows below, we resolve this ambiguity by defining the polarization plane according to a static observer.  {Further details related to the choice of the tetrad and properties of the propagation equation \eqref{eq:master} are discussed in App.~\ref{app:null.tetrads} and App.~\ref{app:shortwave}.}

Since the propagation equations are valid at the order $1/(\omega L)$, where $L$ is the characteristic length scale, we look for a perturbative solution using it as a small dimensionless parameter. The trajectory is thus represented by
\begin{align}
	x^\mu = \accentset{\circ}{x}^\mu + (\omega L)^{-1} x^\mu_{(1)} + \cO\big((\omega L)^{-2}\big),
\end{align}
where $\accentset{\circ}{x}^\mu$ is a solution of the geodesic equation. Therefore, the right-hand side of Eq.~\eqref{eq:master} must include the unperturbed tetrad vectors, $\accentset{\circ}{l}^\mu = \accentset{\circ}{x}^\mu/d\tau$, etc. Consequently, the propagation equations for the tetrad should be enforced only at the zeroth order.\ Hence, Eq.~\eqref{tetrad-prop} reduces to the parallel propagation described by Eq.~\eqref{paral}. As we consider only the unperturbed tetrad, we drop the $^\circ$ label from the vectors to simplify the notation in what follows.

Using Schwarzschild coordinates, a general spherically symmetric metric is given by
\begin{align}
	ds^2 = - f dt^2 + f^{-1} dr^2 +  {r^2} d\Omega_2,
\end{align}
where $f = 1 - r_\sg/r$ and $d\Omega_2 = d\theta^2 + \sin^2 \! \theta \, d\phi^2$. The frequency $\omega$ is the component $l^t$ in the asymptotically flat region and   coincides with the conserved energy $\varepsilon=\hbar \omega$ of a fictitious null point particle. It also sets the scale of the tangent vector and the affine parameter. We rescale them such that $l^t \to 1$ in the asymptotic region.

It is convenient to express distances and other quantities with the dimension of length as ratios of the Schwarzschild radius, $\rho\defeq r/r_\sg$. In spherical symmetry, geodesic motion can be confined to a plane without loss of generality, and we choose the coordinates such that it is identified with the equatorial plane defined by $\theta \equiv \tfrac{\pi}{2}$. Then, the nonzero components of the vector $\al$ are given by
\begin{align}
& l^t \defeq \frac{dt}{d\tau}=\frac{1}{f(\rho)}=\frac{1}{1-\rho^{-1}}, \label{eq:lt}\\
& l^r\equiv p\defeq \frac{dr}{d\tau}=\mp\sqrt{1-\frac{b^2}{\rho^2}f(\rho)}, \\
& l^\phi \defeq \frac{d\phi}{d\tau}=\frac{b}{r_\sg\,\rho^2},
\end{align}
where the $-$ ($+$) sign correspond to the ingoing (outgoing) part of the geodesic trajectory, and $b$ is the reduced impact parameter $b \defeq \ell/ r_\sg$.

While the results are independent of the choice of individual vectors in the polarization plane, a convenient choice of the tetrad vectors significantly simplifies the analysis. The polarization plane is defined by $\bfl$, and we use the Newton gauge of Ref.~\cite{BT:11} to choose individual vectors. In stationary spacetimes, it is defined by the wave vector and the local free-fall acceleration experienced by a static observer. With the above conventions, the three-vector form of the polarization vectors in the Schwarzschild spacetime is given by
\begin{align}
	\bfe_1 = \left(-\frac{b f}{\rho},0,\frac{p}{r_\sg\rho}\right), \quad \bfe_2 = \left(0,\frac{1}{r_\sg\rho},0\right).
\end{align}
The tetrad is completed by setting
\begin{align}
	\an=\frac{1}{2}\left(1,f,0,-\frac{b }{r_\sg\rho^2}\right) ,
\end{align}
and $e^t_j=0$. The Newton gauge has several useful properties. For our purposes, the most important one is that the tetrad vector components remain nonsingular over the entire geodesic trajectory. From a three-dimensional perspective, the geodesic propagation simply rigidly rotates $\mathbf{l}$ and $\bfe_1$ around the constant $\bfe_2$.

At leading order, Eq.~\eqref{tetrad-prop} requires the parallel propagation of the  tetrad vectors $\al$, $\ave_1$, and $\ave_2$.\ However, while $\nabla_\al \al=\nabla_\al\ave_2=0$,
\begin{align}
	\nabla_\al \an &= \sqrt{2}\alpha \ave_1=\alpha(\am+\bar\am), \\
	\nabla_\al \ave_1 &= \sqrt{2}\alpha \al \;\; \Leftrightarrow \;\; \nabla_\al \am=\nabla_\al\bar\am= \alpha \al,
\end{align}
where
\begin{align}
	\alpha = - \frac{1}{2\sqrt{2}r_\sg}\frac{b}{\rho^3} .
\end{align}
Explicit integral expressions for a null frame that is propagated along the null geodesic with the tangent $\al$ for Kerr spacetimes are given in Ref.~\cite{M:83}. Generalizations to higher-dimensional spacetimes are provided in Ref.~\cite{KFKC:09}. However, in our case, it is simpler to perform the frame adjustment explicitly.

The null tetrad transformation of type I \cite{C:92,F:20} [cf.\ Eq.~\eqref{app:eq:A.I}],
\begin{align}
	& \an\to \an_F=\an+a(\am+\bar\am)+a^2\al, \\
	& \am\to\am_F=\am+a\al, \quad \bar\am\to\bar\am_F=\bar\am+a\al,
\end{align}
where the function $a(r)$ satisfies
\begin{align}
	a'(r)p = -\alpha ,
	\label{a-alpha}
\end{align}
{preserves} the parallel transport of the tetrad along the null geodesic \cite{DT:23}. Choosing the initial condition $a(r_\rin)=0$ ensures that the adjusted tetrad is the transformed Newton gauge tetrad. Taking this into account, we write the acceleration as
\begin{align}
	w^\mu = - \frac{\sigma}{\omega} R^\mu_{~\nu\alpha\beta} l^\nu (e_{(1)}^\alpha + al^\alpha) e_{(2)}^\beta.
\end{align}
On the Schwarzschild background, the only nonzero first-order driving term is
\begin{align}
	w^\theta = \frac{\sigma}{\omega r_\sg^3} \left[ \frac{3bp}{2\rho^5}+\frac{a}{\rho^4 f}\left(\frac{1}{2}-p+\frac{b^2}{\rho^2}f\right)\right] .
	\label{force}
\end{align}
Starting with Eq.~\eqref{eq:master}, using $\nabla_\al \al=0$, and keeping only terms linear in $1/(\omega r_\sg)$ results in the linear second-order equations for $x^\mu_{(1)}$. As the appropriate initial conditions are $x^\mu_{(1)}(r_\rin)=0$ and $\dot x^\mu_{(1)}(r_\rin)=0$, only the $\theta$ component (i.e., motion outside of the geodesic plane) has a first-order contribution, in agreement with the results of Refs.~\cite{DMS:19,AO:23}. Setting $\theta_{(1)}=\theta-\pi/2\eqdef\vartheta$, we obtain
\begin{align}
	\frac{d^2\vartheta}{d\tau^2}+\frac{2}{r_\sg\rho}\frac{d\vartheta}{d\tau}p+\frac{ {b}^2}{r_\sg^2\rho^4}\vartheta=w^\theta\big(r(\tau)\big).
	\label{tdom}
\end{align}
The evolution equation~\eqref{tdom} with the acceleration given by Eq.~\eqref{force} and the function $a(r)$ being the solution of Eq.~\eqref{a-alpha} with appropriate boundary conditions form the starting point of our numerical calculations and analytical evaluations.

\begin{figure*}[!htbp]
	\subfloat[{Acceleration $w^\theta(\chi)$ in the deflection scenario.}]{
		\includegraphics[width=.485\linewidth]{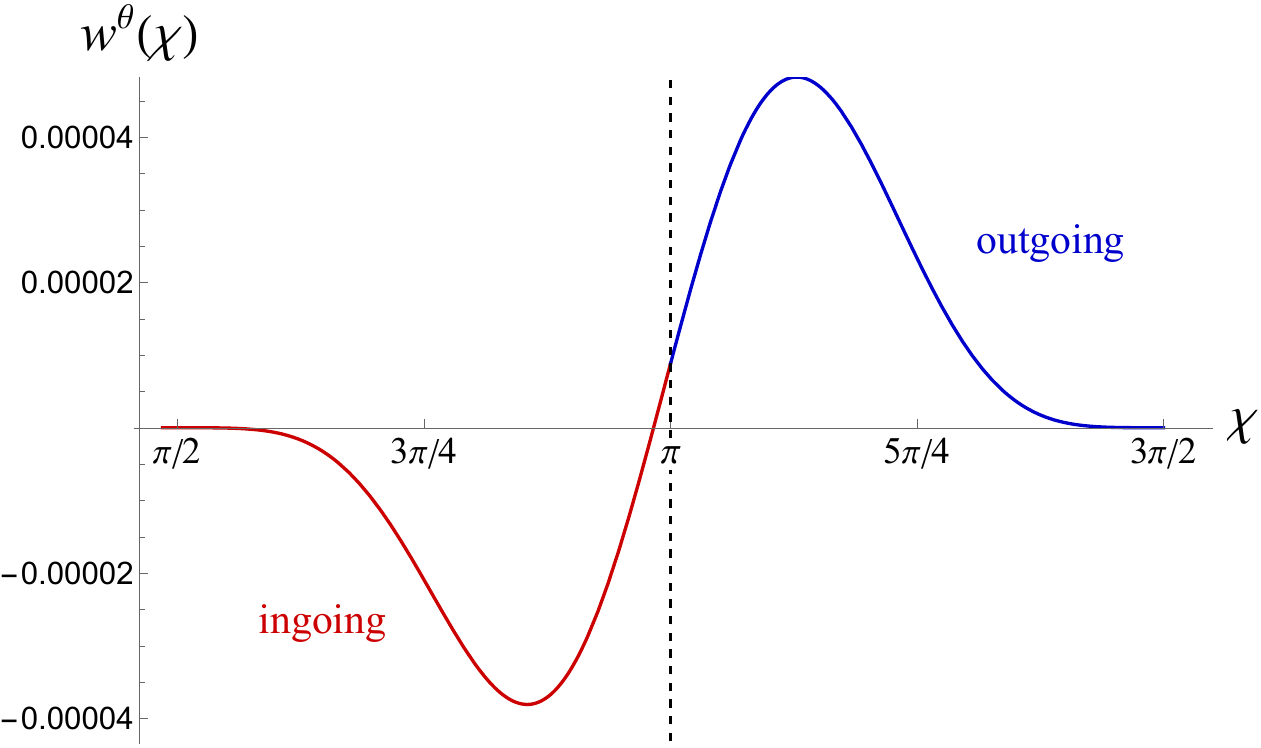}
		\label{fig:w.deflection}
	}\hfill
	\subfloat[{Acceleration $w^\theta(\chi)$ in the emission scenario.}]{
		\includegraphics[width=.485\linewidth]{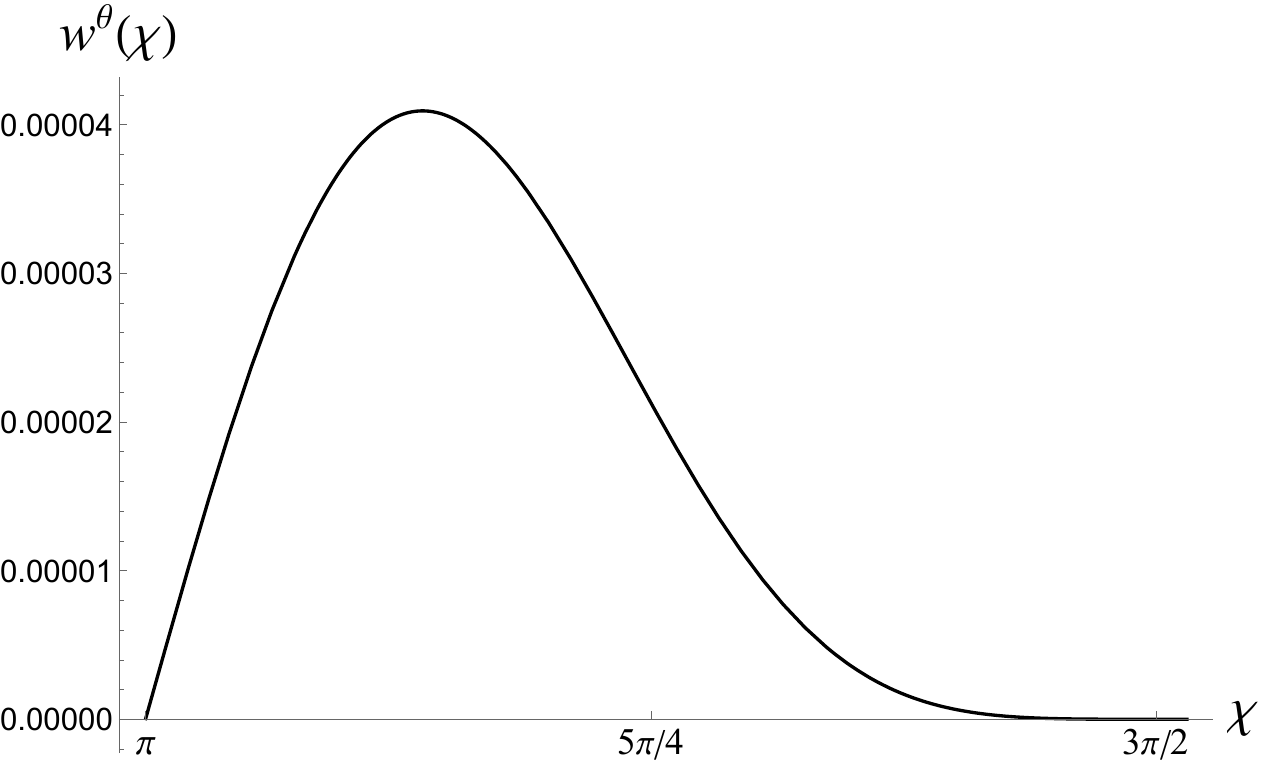}
		\label{fig:w.emission}
	}
	\caption{Acceleration $w^\theta(\chi)$ in the deflection (left) and emission (right) scenario for the parameter choices $r_\sg=\sigma=\omega=1$ and $b_1=10$. In the deflection scenario (left), the perihelion $\rho=b_1$ corresponds to $\chi=\pi$, and ingoing and outgoing parts of the trajectory are illustrated in red and blue, respectively. The limit $\rho\to\infty$ corresponds to $\chi\to \chi_\infty$ and $\chi\to 2\pi-\chi_\infty$, where $\chi_\infty$ is given in Eq.~\eqref{chinf}. In the emission scenarios that we consider $\rho_\rin=b_1$. Note that, as $a(\rho_\rin)=0$, the acceleration in the emission scenario (right) at $\rho(\chi=\pi)=b_1$ equals zero, but this is not the case in the deflection scenario (left).}
	\label{fig:w.deflection.emission}
\end{figure*}

The tetrad transformation parameter $a(r)$ has an explicit form in terms of elliptic integrals,
\begin{align}
	p(\rho) = \mp\sqrt{(\rho-b_1)(\rho-b_2)(\rho-b_3)/\rho^3}.
\end{align}
The three roots of $p(\rho)$ satisfy \cite{C:92}
\begin{align}
	b_1+b_2+b_3=0,
\end{align}
as well as
\begin{align}
	\frac{1}{b_1}+\frac{1}{b_2}+\frac{1}{b_3}=1, \qquad b_1b_2b_3=b^2.
\end{align}
The largest real root $b_1$ is the reduced coordinate radius of the point corresponding to the closest approach of the trajectory to the origin (also referred to as the perihelion). The critical value of the impact parameter $b_\text{cr}=3\sqrt{3}/2$ describes geodesics reaching the light ring. In this article, we only consider scenarios with $b>b_\text{cr}$.

For $b\gg 1$, we have
\begin{align}
	& b_1 = b - \frac{1}{2} - \frac{3}{8b} + \ldots , \\
	& b_2 = - b - \frac{1}{2} + \frac{3}{8b} + \ldots , \\
	& b_3 = 1 + \frac{1}{b^2} + \ldots,
\end{align}
and
\begin{align}
	p=p_0(\rho)+\cO(b_3/b_1), \quad p_0=\mp\sqrt{1-b_1^2/\rho^2}.
	\label{p-appro}
\end{align}
Using this approximation, we find
\begin{align}
	a = \frac{(p_0+1)b}{2\sqrt{2}b_1^2} + \cO(b^{-2}) = \frac{(p_0+1)}{2\sqrt{2}b}+\cO(b^{-2})
\end{align}
for $\rho_\rin=\infty$, and
\begin{align}
	a=\frac{|p_0|}{2\sqrt{2}b}+\cO(b^{-2})
	\label{a-rad}
\end{align}
if the trajectory starts at $\rho_\rin=b_1$. Recalling that $b\gg1$, we find
\begin{align}
	w^\theta=\frac{\sigma}{2\omega r_\sg^3}\frac{ b}{\rho^3}\big(\mathfrak{f}_1+\mathfrak{f}_2+\mathfrak{f}_3\big) ,
\end{align}
where
\begin{align}
	& \mathfrak{f}_1=\frac{3p_0}{\rho^2}\big(1+\cO(b_3/b)\big), \\
	& \mathfrak{f}_2=\frac{(p_0+1)}{\rho^3}\big(1+\cO(b_3/b)\big), \\
	& \mathfrak{f}_3=-\frac{\big(p_0^2-1\big)(p_0+1)}{2b_1^2\rho}\big(1+\cO(b_3/b)\big).
\end{align}
As can be seen in Fig.~\ref{fig:w.deflection}, the change in the signature of $p_0$ at $b_1$ introduces a slight asymmetry in the acceleration $w^\theta(\chi)$. It manifests itself in small differences in some parameters that are described in Sec.~\ref{astro}. The acceleration $w^\theta(\chi)$ for the emission scenario discussed in Sec.~\ref{sec:emission} is shown in Fig~\ref{fig:w.emission}.

\section{Solutions for the deflection \newline and emission of light} \label{solutions}
We consider two scenarios in what follows, namely the deflection of incoming light and the emission of light by a massive body. In the deflection scenario, a light beam originating from infinity approaches the gravitating center at the minimal coordinate radius $\rho=b_1$ and is detected by an observer at some $\rho_\text{O} \gg b_1$. This is the physical setting of classical light deflection tests of general relativity with the light ray grazing the solar radius at $b_1$ \cite{W:18}. The leading polarization-dependent correction describes the deviation from the geodesic propagation plane.

The second scenario deals with the emission of electromagnetic radiation off-center with respect to the line of sight that connects the emitting object and the observer. The mathematical methods used to obtain solutions are the same in both cases. We will describe them based on the deflection scenario in Sec.~\ref{subsec:deflection} and report results for the emission scenario in Sec~\ref{sec:emission}. Additional mathematical details and some of the explicit expressions are presented in App.~\ref{sis} and App.~\ref{app:full.expr}.

\subsection{Deflection of light} \label{subsec:deflection}
\begin{figure}[!htbp]
	\centering
	\includegraphics[width=.95\linewidth]{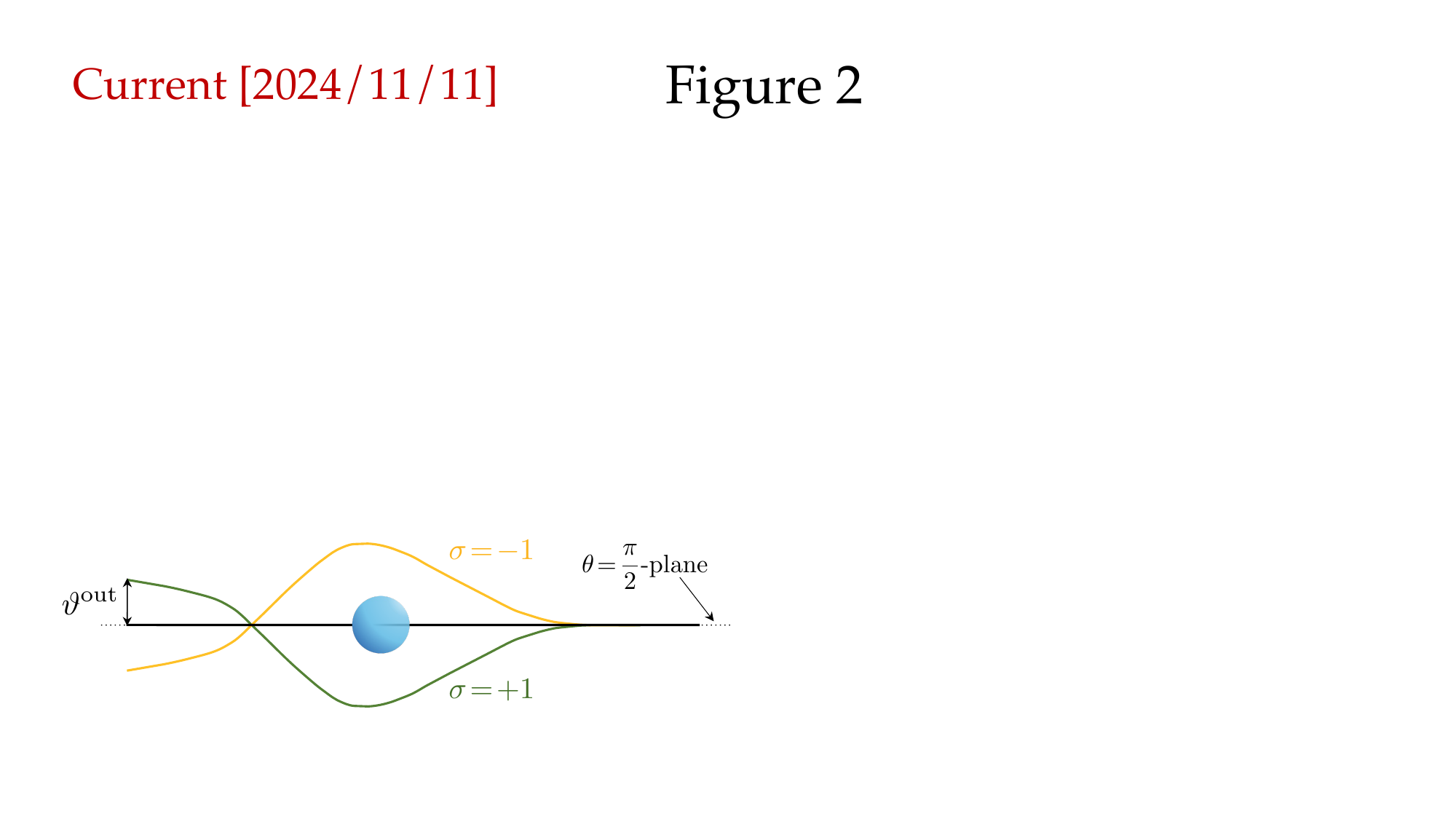}
	\caption{Schematic representation of the polarization-induced deviation for a light ray incoming from the right in the deflection scenario for left $\sigma=-1$ (yellow line) and right $\sigma=+1$ (green line) circularly polarized light. The geodesic trajectory (solid black line) remains in the $\tfrac{\pi}{2}$-plane indicated by the dotted black line. The light rays cross the $\tfrac{\pi}{2}$-plane at $r=r_0$ [cf.\ Eq.~\eqref{dev1.2} and Fig.~\ref{lens2}].}
	\label{fig:schematic.general}
\end{figure}

While Eq.~\eqref{tdom} does not have a solution with a closed analytic form, it can be solved numerically using a convenient parametrization of the geodesic trajectory (see App.~\ref{schw} for details).\ Figure~\ref{fig:schematic.general} schematically depicts the evolution of the polarization-dependent deviation from the geodesic plane for left (yellow) and right (green) circularly polarized light. At $r= r_0$, which is given explicitly in Eq.~\eqref{dev1.2} below, the light rays cross the plane of the geodesic propagation.

Using the linearity of Eq.~\eqref{tdom}, it is possible to obtain its analytic solution iteratively. A brief discussion of our iterative procedure is presented in App.~\ref{sis}. We set
\begin{align}
	\vartheta=\sum_{j=0}^{N}\vartheta_j +\vartheta_{N+1}\defeq S_N+\vartheta_{N+1},
	\label{eq:vartheta.partialsum.decomposition}
\end{align}
where $\vartheta_j$ with $j=0,\ldots, N$ denote solutions of the equations
\begin{align}
	\frac{d^2\vartheta_j}{d\tau^2}+\frac{2}{r_\sg\rho}\frac{d\vartheta_j}{d\tau}p_0=w_j ,
	\label{2iter}
\end{align}
and we have replaced the function $p$ by its approximation Eq.~\eqref{p-appro} since we are interested in the leading order in $b$. If needed, expressions that include subleading terms can be obtained by the same method presented below after $a(r)$ and $p(r)$ are expanded to the desired order.

In this sequence of equations $w_0\defeq w^\theta$, and for $1 \leqslant j \leqslant N$ the preceding term in the series drives the subsequent equation via $w_j \defeq -b^2 \vartheta_{j-1}/(r_\sg^2\rho^4)$.\ The term $\vartheta_{N+1}$ is the remainder term that makes the decomposition exact. It satisfies the original equation with the right-hand side given by $w_{N+1}$.

\begin{figure*}[!htbp]
	\subfloat[{Polarization-induced corrections $\vartheta$ in the deflection scenario.}]{
		\includegraphics[width=.485\linewidth]{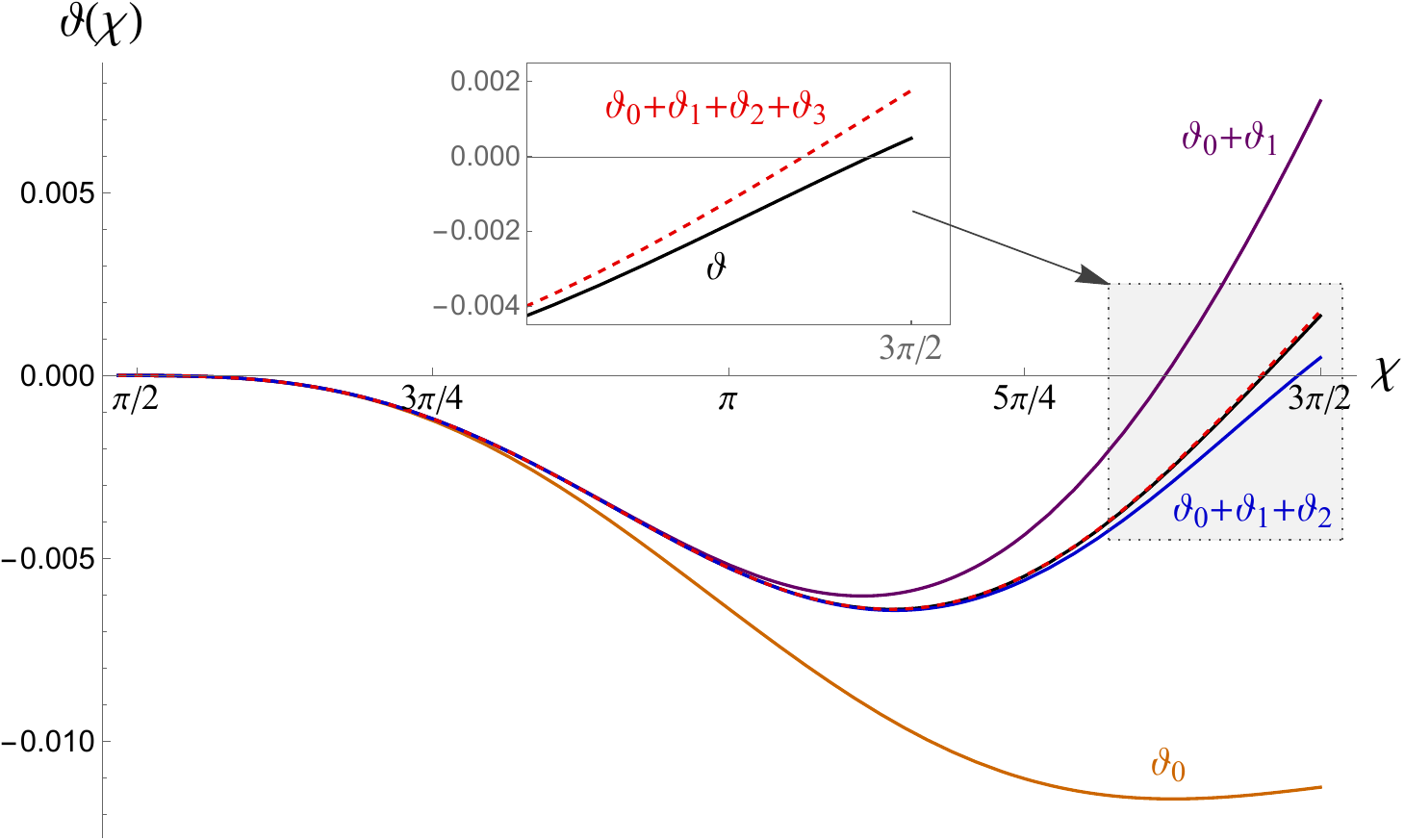}
		\label{vartheta_scattering}
	}\hfill
	\subfloat[Polarization-induced corrections $\vartheta$ in the emission scenario.]{
		\includegraphics[width=.485\linewidth]{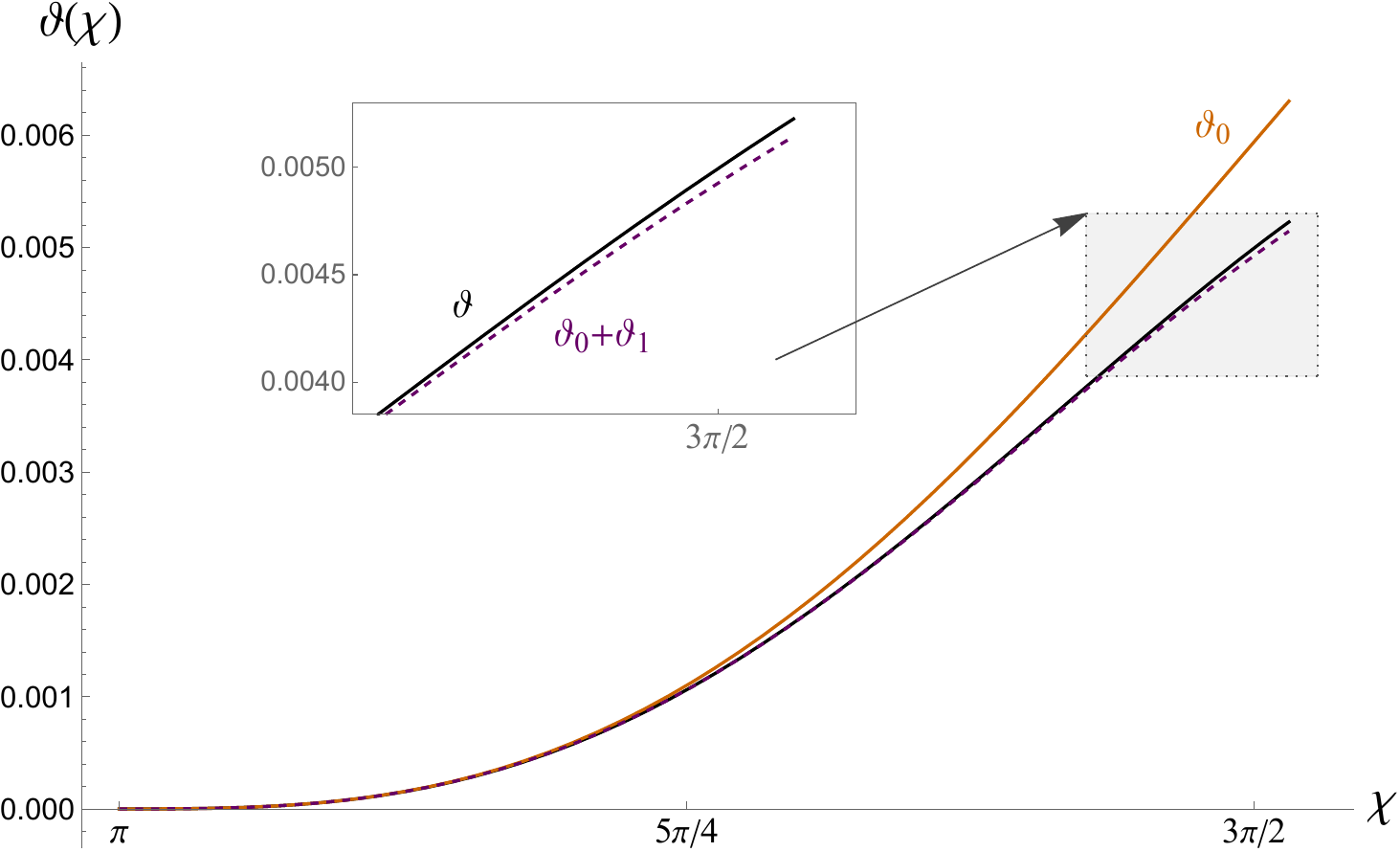}
		\label{vartheta_emission}
	}
	\caption{Polarization-dependent corrections $\vartheta(\chi)$ in the deflection (left) and emission (right) scenario for the parameter choices $r_\sg=\sigma=\omega=1$ and $b_1=10$. In both cases, the numerical solution is shown as a solid black line labeled as $\vartheta$. The iteratively obtained analytical solutions are illustrated up to $S_3$ (left) and $S_1$ (right) and are labeled accordingly [cf.\ Eq.~\eqref{eq:vartheta.partialsum.decomposition}].}
	\label{fig:corr.deflection.emission}
\end{figure*}

The advantage of this approach is that for each $j<N+1$ it is possible to obtain an analytic solution. In fact, the sequence of partial sums $\sum_{j=0}^N\vartheta_j$ is convergent (see App.~\ref{sis}) and only a relatively small number of terms is needed to attain good agreement with numerical calculations (see Fig.~\ref{fig:corr.deflection.emission}).

Equation~\eqref{2iter} is transformed into the first-order linear ordinary differential equation (ODE) by introducing
\begin{align}
	q_j = \frac{d\vartheta_j}{d\tau}=\vartheta'_j \frac{p}{r_\sg},
	\label{defqj}
\end{align}
where $\vartheta_j' \defeq d\vartheta_j/d\rho$.\ Thus the leading-order equation is represented by the first order ODE
\begin{align}
	q'_jp_0+\frac{2}{\rho}q_jp_0=r_\sg w_j ,
\end{align}
which makes exact analytic solutions possible. Using $\rho$ as the evolution parameter necessitates the combination of solutions along the ingoing and outgoing parts of the geodesic trajectory. The initial condition for the ingoing part is $q^\rin_j (\infty)=0$, and the outgoing and ingoing parts of the solution are matched by setting $q_j^\rou(b_1)=q^\rin_j(b_1)$.

The ingoing part of the solution is given by
\begin{align}
	q^\rin_j(\rho)=\frac{r_\sg}{\rho^2}\left(\int_\rho^\infty x^2\frac{w_j^\rin(x)}{|p_0(x)|}dx+c_j^\rin\right), \label{q-in}
\end{align}
where the constants $c^\rin_j$ are set to zero as $q\equiv 0$ in the absence of the gravitational spin Hall effect. The deflection is obtained by integrating Eq.~\eqref{defqj},
\begin{align}
	\vartheta^\rin_j(\rho)=r_\sg \int_\rho^\infty\frac{ q_j^\rin{(x)}}{|p_0{(x)}|}dx, \label{theta-in}
\end{align}
where we used the initial condition $\vartheta^\rin(\infty)=0$.

For the outgoing segment of the trajectory, we have
\begin{align}
	q^\rou_j(\rho)=\frac{r_\sg}{\rho^2}\left(\int_{b_1}^\rho x^2\frac{w_j^\rou(x)}{|p_0(x)|}dx\right) + q_j^\rin(b_1)
	\label{q-out}
\end{align}
and
\begin{align}
	\vartheta^\rou_j(\rho) = r_\sg\int_{b_1}^\rho \frac{ q_j^{\rou}{(x)}}{|p_0{(x)}|} dx + \vartheta^\rin_j(b_1) .
	\label{theta-out}
\end{align}
The explicit construction of $\vartheta_0$ is detailed in App.~\ref{sis}. Full expressions up to the seventh order are provided in the \textsc{GitHub} repository listed as Ref.~\cite{GitHubRep}.

A comparison of the numerically obtained $\vartheta$ with the first three iterations $S_3 = \vartheta_0 + \vartheta_1 + \vartheta_2 + \vartheta_3$ [cf.\ Eq.~\eqref{eq:vartheta.partialsum.decomposition}] in the deflection scenario is shown in Fig.~\ref{vartheta_scattering}. The partial sums quickly converge to $\vartheta$, with $N=3$ illustrated by the dashed red line already producing a very good match.

Using the iterative solution, it is possible to obtain several key characteristics of the gravitational spin Hall effect, namely the deflection at the perihelion, the radius of the recrossing of the equatorial plane, and the asymptotic deflection, which are given by
\begin{alignat}{2}
	&\vartheta(b_1)=-\gb\frac{\sigma}{\omega r_\sg b^2}, & \qquad & \gb \approx0.50,
	\label{dev1.1}
	\\
	&  r_0=\rho_0r_\sg=\gd  b^2 r_\sg,  & \qquad & \gd \approx 0.56,
	\label{dev1.2}
	\\
	&\vartheta^\rou(\infty)=\gc_\infty \frac{\sigma}{\omega r_\sg b^3}, & \qquad & \gc_\infty \approx 1.78 ,
	\label{dev1.3}
\end{alignat}
respectively. The full analytical expressions for $\gb$, $\gd$, and $\gc_\infty$ are provided in App.~\ref{app:full.expr}. They correspond to the leading-order expressions for $b_1 = b + \cO(1)$. The next-order terms involve contributions from terms that are linear in $b_3$ in the expansion of $p(\rho)$.

\subsection{Emission of light} \label{sec:emission}
Any light ray with nonzero impact parameter $b \neq 0$ in the Schwarzschild spacetime defines a plane. For $b=0$ on the other hand, the source, the central mass $r_\sg/2$, and the observer are colinear, and no such plane is defined. The Newton gauge cannot be introduced, although the existence of polarization-dependent deviations would have given it an absolute meaning in this case. The absence of such deviations can also be inferred from Eq.~\eqref{force} with $b=a=0$ corresponding to the outgoing radial null geodesic. Taking into account that the tetrad rotation parameter is given by Eq.~\eqref{a-rad}, the analysis of the emission scenario is analogous to the deflection scenario discussed in Sec.~\ref{subsec:deflection}.

\begin{figure*}[!htbp]
	\centering
	\includegraphics[width=.93\linewidth]{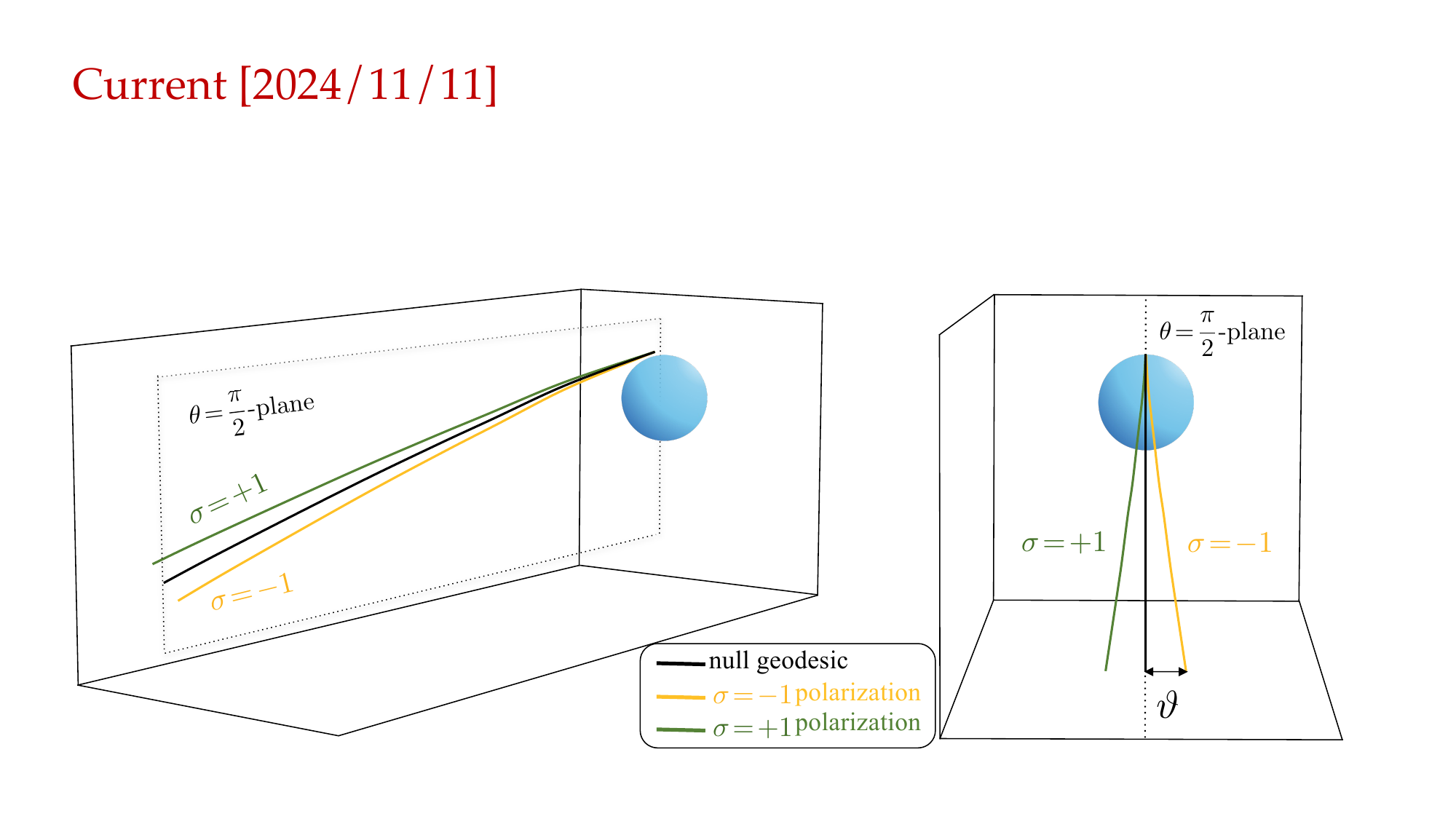}
	\caption{Schematic representation of the polarization-induced deviations $\vartheta$ for nonradially emitted outward propagating light rays. As in Fig.~\ref{fig:schematic.general}, the $\tfrac{\pi}{2}$-plane is indicated by the dotted black line(s), the geodesic trajectory corresponds to the solid black line, and left and right circularly polarized light rays are indicated by the yellow and green lines, respectively.}
	\label{fig:schematic.nonradial}
\end{figure*}

Here, we consider trajectories with $\rho_\rin = b_1$. The general case $0 < b_1 < \rho_\rin$ is treated similarly. A typical trajectory is depicted schematically in Fig.~\ref{fig:schematic.nonradial}. Unlike the deflection scenario, the magnitude of the deviation from the geodesic plane always increases and there is no reconvergence. The asymptotic deflection from the geodesic plane is given by
\begin{align}
	\vartheta^\rou(\infty)=\gc \frac{\sigma}{\omega r_\sg {b^2}}, \qquad \gc \approx 0.494 .
	\label{dev2}
\end{align}
This expression is valid for $b_1 \gg r_\sg$. The deviations $\vartheta^\rou(\infty)$ [Eq.~\eqref{dev2}] in the emission scenario and $\vartheta(b_1)$ [Eq.~\eqref{dev1.1}] in the deflection scenario are of the same order $1/b^2$ and match very closely, i.e., up to about $99 \%$. The $1 \%$ discrepancy is due to the slight asymmetry in $w^\theta$ in the deflection scenario (see Fig.~\ref{fig:w.deflection}). The full analytical expression for $\gc$ is provided in App.~\ref{app:full.expr}.

A comparison of the numerically obtained $\vartheta$ with the first iteration $S_1 = \vartheta_0 + \vartheta_1$ [cf.\ Eq.~\eqref{eq:vartheta.partialsum.decomposition}] in the emission scenario is shown in Fig.~\ref{vartheta_emission}. Once again, the partial sums quickly converge to $\vartheta$, with $N=1$ illustrated by the dashed purple line already producing a very good match.

\section{Observational consequences} \label{astro}
In this section, we discuss the effects of polarization-induced deviations from the geodesic motion of photons and apply our scaling relations to observationally relevant astrophysical scenarios.  {By adjusting the parameter $\sigma$ in Eq.~\eqref{eq:master} appropriately, the formalism presented in this article can be used to treat massless fields of a different spin (e.g., spin-\textonehalf\ Dirac fields and spin-2 fields in linearized gravity) analogously \cite{AO:23,OK:23}.} Numerical investigations of time delay effects, additional features resulting from the angular momentum of the gravitating body, as well as implications for gravitational waves are studied in Refs.~\cite{OSZ:23,AO:23,FS:24}.

\subsection{Gravitational lensing}
Gravitational lensing encompasses all effects of gravitational fields on the propagation of electromagnetic waves. Hence, the first classical test of general relativity \cite{W:18} is also the first observation of gravitational lensing \cite{W:98,P:04}. Initially regarded merely as a geometric curiosity \cite{W:98}, gravitational lensing established its usefulness in astrophysics by making visible multiple quasar images, elongated arcs of distant galaxies, and rings of extragalactic radio sources \cite{W:98,SEF:92}. Nowadays, it is mostly used in the detection of extrasolar planets, observations constraining the distribution of dark matter, and the evaluation of cosmological parameters \cite{P:04,CK:18}. Interestingly, it also played a role in the generation of physically accurate visual effects for the movie \textit{Interstellar} \cite{JTFT:15}.

The analysis of gravitational lensing is usually performed using the geometric optics approximation \cite{SEF:92,W:98,P:04}. While wave optics is used in the evaluation of the brightness of images as well as their magnification \cite{SEF:92,P:04}, the use of geometric optics is sufficient in all other cases. The polarization of electromagnetic waves was discovered in the analysis of gravitational lensing observations \cite{EHT:23,CK:18,P:14,B:22}, but the effects of birefringence are typically neglected.

\begin{figure}[!htbp]
	\centering
	\includegraphics[width=.95\linewidth]{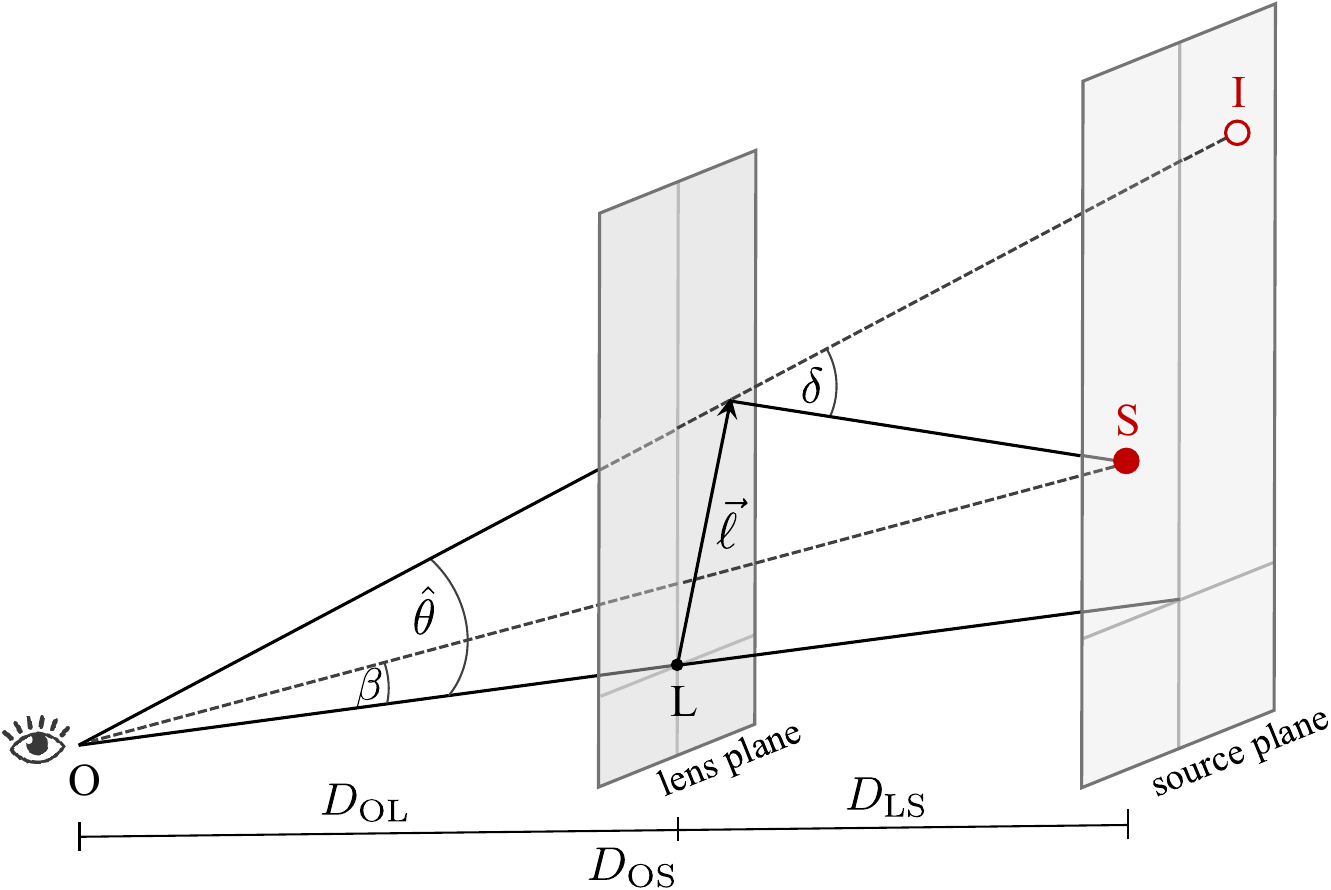}
	\caption{Schematic illustration of the gravitational lens geometry in the thin point-particle lens approximation. The source S is located at a distance $D_\text{OS}$ from the observer and appears to the observer O as the image I. The distances between the lens and the source and the lens and the observer are indicated by $D_\text{LS}$ and $D_\text{OL}$, respectively. On an approximately flat background, the distances are related via $D_\text{OS} = D_\text{OL} + D_\text{LS}$. When the large scale structure of the universe is taken into account, the appropriate generalizations of the Euclidean quantities must be used \cite{SEF:92,W:98,CK:18}.}
	\label{lens1}
\end{figure}

Time and time again, technological developments have turned previously unthinkably futuristic measurements first into cutting edge observations, and then into standard research tools. Our scaling estimates will help in evaluating the significance of birefringence effects for various gravitational lensing scenarios. We consider the simplest model of gravitational lensing --- a thin point-particle lens \cite{SEF:92}. While many realistic scenarios are much more involved, the resulting analytic expressions provide quick estimates of basic relations, such as that between the lens mass and the angular separation of images. The thin point-particle lens model describes the effect of the gravitational field on the propagation of electromagnetic waves via approximate tracing of null geodesics in the Schwarzschild spacetime with mass $M=r_\sg/2$. For $\ell \gg r_\sg$, the actual trajectory can be approximated by its two asymptotes. The setup is shown schematically  in Fig.~\ref{lens1}.

The deviation from the geodesic trajectory occurs as a consequence of gravitational deflection, which is given by $\delta=2b$ in the thin point-particle lens approximation. Without the deflection, the observed angular separation between the point mass and the light ray would be $\beta$, while the actual observed separation is then given by $\hat\theta=\ell/D_\text{OL}$.

Based on the geometry of Fig.~\ref{lens1}, this results in
\begin{align}
	\beta=b\frac{r_\sg }{D_\text{OL}} -\frac{2}{b}\frac{D_\text{LS}}{D_\text{OS}} .
	\label{gl-main}
\end{align}
Introducing the characteristic angle
\begin{align}
	\alpha_0=\sqrt{2r_\sg\frac{D_\text{LS}}{D_\text{OS}D_\text{OL}}}
\end{align}
allows us to rewrite this equation as
\begin{align}
	\hat\theta^2-\beta\hat\theta-\alpha_0^2=0 .
	\label{ang-L}
\end{align}
For a point-like lens, the two real roots of Eq.~\eqref{ang-L} correspond to the two images of the source with angular separation
\begin{align}
	\Delta\hat\theta=\sqrt{4\alpha_0^2+\beta^2}>2\alpha_0.
\end{align}
The two images are of comparable brightness only if the angular separation between the
deflecting mass and the image $\hat\theta$ is of the order of the characteristic angle $\alpha_0$. We now investigative how polarization-induced birefringence affects imaging in this case.

In a typical scenario that is commonly considered \cite{SEF:92,W:98}, the source is situated much further from the gravitational lens than the observer, i.e., $D_\text{OS} \sim D_\text{LS} \;  {\gg D_\text{OL}}$, and the condition $\hat\theta\sim\alpha_0$ implies
\begin{align}
	b \sim \sqrt{\frac{D_\text{OL}}{r_\sg}} \gg 1
\end{align}
and
\begin{align}
	\vartheta \sim \frac{1}{\omega r_\sg} \left(\frac{r_\sg}{D_\text{OL}}\right)^{3/2} \!\! \sim \frac{\alpha_0}{\omega D_\text{OL}}\ll\alpha_0,
\end{align}
where we have assumed that the plane crossing satisfies $r_0 \ll D_\text{OL}$.

\begin{figure*}[!htbp]
	\centering
	\includegraphics[width=.95\linewidth]{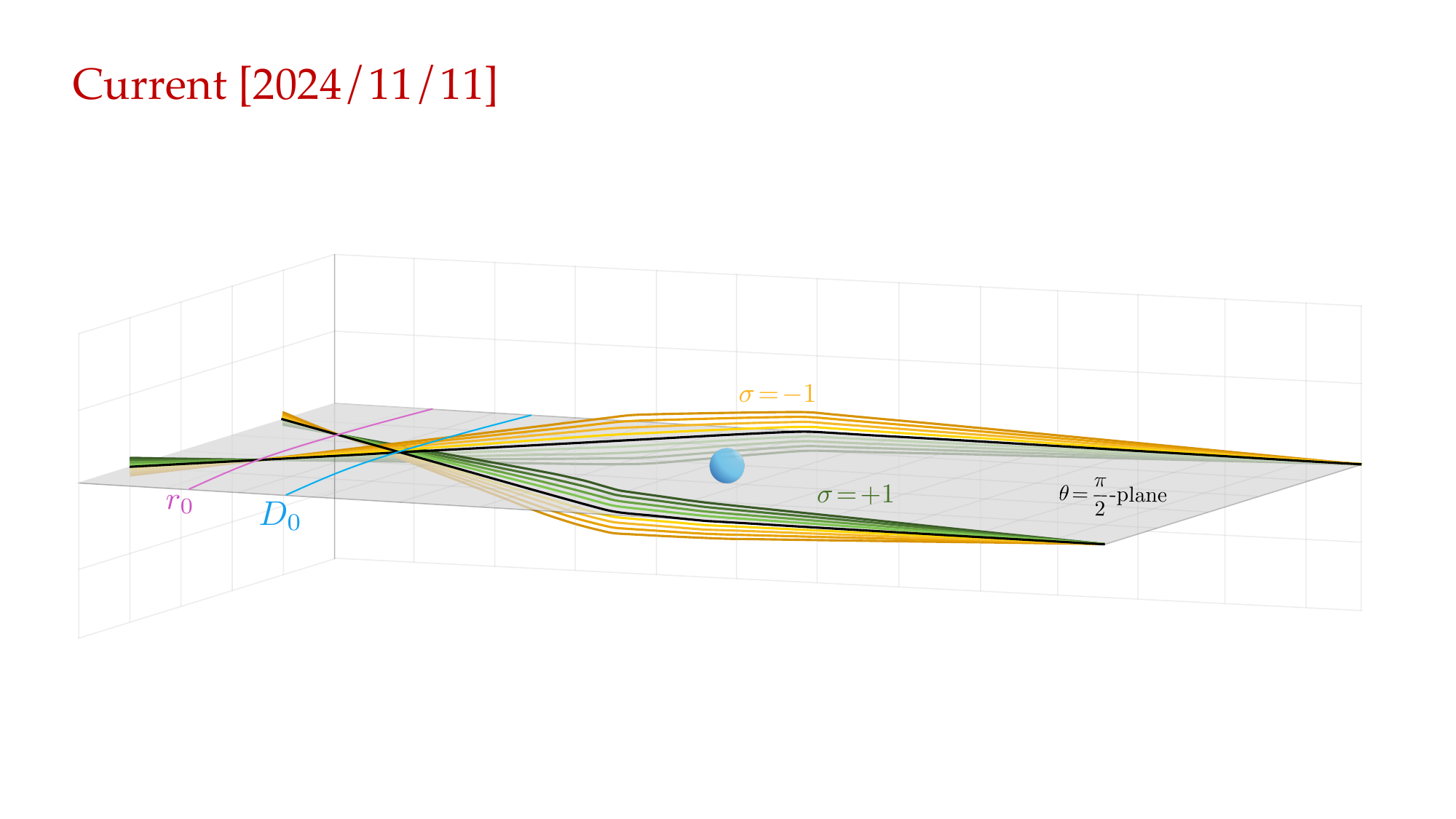}
	\vspace*{-3mm}
	\caption{Schematic representation of gravitationally induced birefringence in the solar gravity lens setting. The evolution of two geodesic rays (indicated by the solid black lines) from the pencil of rays determines the equatorial $\tfrac{\pi}{2}$-plane (shaded in gray). The geodesics converge at $D_0$. Light rays with right circular polarization ($\sigma=+1$, indicated in various shades of green) and left circular polarization ($\sigma=-1$, indicated in various shades of yellow/orange) exit the equatorial plane and return to it at $r_0$, continuing on the other side. Figure~\ref{fig:schematic.general} provides a side view of this scenario.}
	\label{lens2}
\end{figure*}

Table~\ref{Table1} presents the deviations $\vartheta(b_1)$ and $\vartheta^\rou(\text{Earth})$ for visible red light with frequency $\omega_{\text{(red)}} = 400$ THz and the radio frequency $\omega_{\text{(radio)}} = 15$ MHz. The beams of light are incoming from $\infty$ and are deflected by the Sun, Proxima Centauri (our closest star), and RX J1856.5-3754 (our closest neutron star), respectively, before being observed on Earth. For $\omega_{\text{(red)}}$, the polarization-induced corrections are negligible. For $\omega_\text{(radio)}$ on the other hand, the deviations are significantly larger, most notably the deviation $\vartheta^\rou_\text{(radio)} \text{(Earth)} = \pm 2.4 \times 10^{-6}$ rad for light deflected onto the Earth by RX J1856.5-3754 (highlighted by the box in the lower right-hand side corner of Tab.~\ref{Table1}). Detecting a deviation of this magnitude requires a radio telescope capable of capturing the 15 MHz radio frequency with an angular resolution of $4.8 \times 10^{-6}$ rad. The radio telescope LOFAR, for instance, is capable of capturing a 15 MHz radio frequency with an angular resolution of $1.5\times 10^{-5}$ rad \cite{LOFAR}. We thus expect that the effect of polarization-induced birefringence may be detected in observations in the near future.

\begin{table*}[!htpb]
	\centering
	\resizebox{\linewidth}{!}{
		\begin{tabular*}{\linewidth}{@{\extracolsep{\fill}}cccc cc cc}
			\toprule \toprule
			\small{Object} & \small{Mass} $[\textup{M}_\odot]$ & \small{Radius} $[r_g]$& \small{Distance} & \small{$\vartheta_\text{(red)}(b_1) [\text{rad}]$} & \small{$\vartheta_\text{(radio)}(b_1) [\text{rad}]$}& \small{$\vartheta^\rou_\text{(red)} \text{(Earth)} [\text{rad}]$} & \small{$\vartheta^\rou_\text{(radio)} \text{(Earth)} [\text{rad}]$}
			\\
			\midrule \\[-3mm]
			\small{Sun} & \small{1.0} & \small{235728} & \small{1 AU} & \small{$\mp 3.6 \times 10^{-22}$} & \small{$\mp 9.7 \times 10^{-15}$} & \small{$\mp 3.4 \times 10^{-24}$} & \small{$\mp 9.0 \times 10^{-17}$}  \\[1ex]
			\small{Proxima Centauri} & \small{0.12} & \small{297422} & \small{4.2 ly} & \small{$\mp 1.9 \times 10^{-21}$} & \small{$\mp 5.0 \times 10^{-14}$} & \small{$\pm 2.3 \times 10^{-26}$} & \small{$\pm 6.0 \times 10^{-19}$}  \\[1ex]
			\small{RX J1856.5-3754} & \small{0.90} & \small{10} & \small{400 ly} & \small{$\mp 2.3 \times 10^{-13}$} & \small{$\mp 6.1 \times 10^{-6}$} & \small{$\pm 9.0 \times 10^{-14}$} & \small{$\boxed{\pm 2.4 \times 10^{-6}}$} \\[1mm]
			\bottomrule \bottomrule
		\end{tabular*}
	}
	\caption{Polarization-induced deviations for light that is approaching from $\infty$ and then deflected by the Sun, Proxima Centauri, and RX J1856.5-3754 towards the Earth. The deviations listed with the ``red'' and ``radio'' subscripts correspond to light frequencies of 400 THz and 15 MHz, respectively. The upper (lower) signature corresponds right (left) circularly polarized light. For convenience, the mass is expressed in terms of solar masses $[\textup{M}_\odot]$, the radius in terms of the Schwarzschild radius $[r_g]$, the distance either in light-years $[\text{ly}]$ or astronomical units $[\text{AU}]$, and polarization-induced deviations in terms of radians $[\text{rad}]$.}
	\label{Table1}
\end{table*}

Similar conclusions are drawn if all distances $D_\text{OL}$, $D_\text{OS}$, and $D_\text{LS}$ are on the same scale, i.e., $D_\text{OL} \sim D_\text{OS}$. On the other hand, if the source is located much closer to the lens than the observer, then
\begin{align}
	b \sim \sqrt{\frac{D_\text{LS}}{r_\sg}}
\end{align}
and
\begin{align}
	\vartheta\sim \frac{1}{\omega r_\sg}\left(\frac{r_\sg}{D_\text{LS}}\right)^{3/2} \!\! \sim \; \frac{D_\text{OL}}{\omega D_\text{LS}^2} \alpha_0 .
\end{align}
In this case, it is conceivable that the birefringence effect is much greater, particularly for long wavelengths.

Table~\ref{Table2} presents an overview of the polarization-induced deviations for light originating at $2b_1$ (i.e., very close to the lens) and deflected towards the Earth. The deviations $\vartheta^\rou\text{(Earth)}$ for light deflected by the Sun and Proxima Centauri are significantly greater. As the source gets closer to the perihelion $b_1$, so does the reconvergence radius $r_0$.

\begin{table*}[!htpb]
	\begin{minipage}{0.85\linewidth}
	\centering
	\begin{tabular*}{\linewidth}{@{\extracolsep{\fill}}c cc cc}
 		\toprule \toprule
 		\small{Object} & \small{$\vartheta_\text{(red)}(b_1) [\text{rad}]$} & \small{$\vartheta_\text{(radio)}(b_1) [\text{rad}]$}& \small{$\vartheta^\rou_\text{(red)}\text{(Earth)} [\text{rad}]$} & \small{$\vartheta^\rou_\text{(radio)}\text{(Earth)} [\text{rad}]$} \\
 		\midrule \\[-3mm]
  		\small{Sun} & \small{$\mp 2.4 \times 10^{-22}$} & \small{$\mp 6.3 \times 10^{-15}$} & \small{$\pm 4.3 \times 10^{-23}$} & \small{$\pm 1.2 \times 10^{-15}$}  \\[1ex]
 		\small{Proxima Centauri} & \small{$\mp 1.2 \times 10^{-21}$} & \small{$\mp 3.2 \times 10^{-14}$} & \small{$\pm 2.3 \times 10^{-22}$} & \small{$\pm 6.2 \times 10^{-15}$}  \\[1ex]
 		\small{RX J1856.5-3754} & \small{$\mp 1.5 \times 10^{-13}$} & \small{$\mp 4.1 \times 10^{-6}$} & \small{$\pm 1.1 \times 10^{-13}$} & \small{$\boxed{\pm 2.9 \times 10^{-6}}$} \\[1mm]
 		\bottomrule \bottomrule
	\end{tabular*}
	\caption{Polarization-induced deviations for light with frequencies of 400 THz (``red'') and 15 MHz (``radio'') originating at $2b_1$ (i.e., very close to the lens) and deflected towards the Earth in radians $[\text{rad}]$. The mass, radius, and distance of the objects under consideration are provided in Tab.~\ref{Table1}.}
	\label{Table2}
	\end{minipage}
\end{table*}

It is worth noting that gravitational lenses exhibit strong spherical aberrations to the extent that there may no longer be a single focal point and the focal length is undefined. For a parallel pencil of rays (whose propagation direction defines the optical axis), $\beta \to 0$ and $D_\text{OS} \approx D_\text{LS} \to \infty$. Equation~\eqref{gl-main} then shows that the bending angle is inversely proportional to the impact parameter, $\ell \approx b_1 r_\sg$, and the locus of points
\begin{align}
	D_0(b) = \frac{\ell^2}{2r_\sg} = \frac{b^2r_\sg}{2}
\end{align}
forms a semi-infinite focal line. For the same value of $b$, it is about $12\%$ closer to the center than the radius of reconvergence $r_0$, as illustrated in Fig.~\ref{lens2}.

A futuristic proposal aims to use such a configuration with the Sun as a lens \cite{E:79}. For a beam that grazes the Sun at $\ell=6.96\times 10^8$ ($r_\sg=2.96\times 10^3$ m), the focal point is situated at $D_\text{OL}=8.19\times 10^{13}$ m, which corresponds to approximately 548 AU, or roughly three light days. Interestingly, the radius of reconvergence is located at $r_0=8.99\times 10^{13}$ m, which is about 613 AU. This explains why there is no change in signature between $\vartheta(b_1)$ and $\vartheta^\rou \text{(Earth)}$ for light deflected by the Sun (see Tab.~\ref{Table1}).

The investigation of the optical properties of this so-called solar gravity lens (SGL) has attracted considerable efforts, including technical characteristics of the probe as well as its positioning in deep space and communications with it \cite{TT:17,TT:21,NASA:20,ES:22,TT:23}. The anticipated results of the SGL are quite remarkable: A probe with a 1-m telescope in the SGL focal region is expected to produce direct high-resolution images of exoplanets, offering a maximum light amplification on the order of $10^{11}$ and an angular resolution of $10^{-11}$ arcsec or $4.9 \times 10^{17}$ rad for a wavelength of $1$ µm. In light of these very ambitious goals, it is worth checking to what extent our predictions may be affected by polarization-dependent birefringence. For light of $1$ µm wavelength approaching from $\infty$ and subsequently deflected by the Sun onto the telescope at the focal point $D_0$ of the gravitational lens, we obtain polarization-induced deviations of $\vartheta^\rou(b_1) = \mp 4.9 \times 10^{-22}$ rad and $\vartheta^\rou\text{(SGL)} = \mp 9.4 \times 10^{-28}$ rad, which is well beyond the angular resolution capabilities of the proposed SGL telescope.

\subsection{Emission of light}
The polarization-induced corrections $\vartheta^\rou$ accumulated in the emission scenario by electromagnetic waves originating at the perihelion $b_1$ and detected by an observer on Earth are summarized in Tab.~\ref{tab:deviations.emission}. Compared to the previously considered scenarios, all deviations $\vartheta^\rou$ are at least one order of magnitude greater.\ An angular resolution of $1.3 \times 10^{-5}$ rad would be required to observe the $\pm 6.4 \times 10^{-6}$ rad deviation created by the neutron star RX J1856.5-3754. This is very close to the angular resolution $1.5 \times 10^{-5}$ rad (at $15$ MHz) of the LOFAR telescope \cite{LOFAR}.

\begin{table}[!htpb]
	\begin{minipage}{0.95\linewidth}
	\centering
		\begin{tabular*}{\linewidth}{@{\extracolsep{\fill}}c cc}
			\toprule \toprule
			\small{Object} & \small{$\vartheta^\rou_\text{(red)}\text{(Earth)} [\text{rad}]$} & \small{$\vartheta^\rou_\text{(radio)}\text{(Earth)} [\text{rad}]$} \\
			\midrule \\[-3mm]
			\small{Sun} & \small{$\pm 3.6 \times 10^{-22}$} & \small{$\pm 9.6 \times 10^{-15}$}  \\[1ex]
			\small{Proxima Centauri} & \small{$\pm 1.9 \times 10^{-21}$} & \small{$\pm 4.9 \times 10^{-14}$}  \\[1ex]
			\small{RX J1856.5-3754} & \small{$\pm 2.4 \times 10^{-13}$} & \small{$\boxed{\pm 6.4 \times 10^{-6}}$} \\[1mm]
			\bottomrule \bottomrule
		\end{tabular*}
		\caption{Polarization-induced deviations for the emission of electromagnetic waves originating at the perihelion $b_1$ and detected on Earth. The mass, radius, and distance of the objects under consideration are provided in Tab.~\ref{Table1}.}
		\label{tab:deviations.emission}
	\end{minipage}
\end{table}

Our numerical calculations suggest that for light propagating close to black holes, the polarization-induced deviations are several orders larger and fall well within the angular resolution capabilities of LOFAR. A comprehensive understanding of the black hole case requires an in-depth theoretical study that will be presented elsewhere.

\section{Discussion} \label{sec:discussion}
We have obtained analytical estimates of gravitational birefringence in spherically symmetric spacetimes for light propagating sufficiently far outside of the Schwarzschild radius of the gravitating object. Two useful extensions of this work naturally present themselves: First, we can obtain estimates for cases of extreme gravitational lensing, e.g., when the object is close to the light ring of a black hole, $D_\text{LS} \simeq \tfrac{3}{2} r_\sg$. This will require accounting for terms of order $b_3/b_1$ or higher. Second, interesting results can be expected to be found in axially symmetric spacetimes.

So far, gravitational birefringence has not been observed experimentally \cite{AO:23}. In the scenarios we have considered here, the effects are too small to be detectable with current technology, although in some cases (with the most significant deviations enclosed by the box in the lower right-hand side corner of Tabs.~\ref{Table1} and \ref{Table2}), they are not far off. In addition, gravitational wave measurements may provide an opportunity for detecting polarization-induced birefringence effects due to their sensitivity at lower frequencies \cite{FS:24,AO:23}. The extreme lensing regime, where birefringence effects may manifest themselves in both the polarization dependence of images and/or properties of astrophysical black hole shadows, is another promising domain \cite{PT:22}. We plan to explore these directions in future work.

\acknowledgements
SM is supported by the Quantum Gravity Unit of the Okinawa Institute of Science and Technology (OIST).	RV is supported by a Macquarie University Research Excellence Scholarship. We would like to thank Joanne Dawson, Elizabeth Cappellazzo,  and Justin Tzou for useful discussions and helpful comments.

\appendix

\section{Properties of null tetrads} \label{app:null.tetrads}
Here, we summarize the relevant properties of a null tetrad that is adapted to a congruence of null curves identified with the trajectories of fictitious classical photons. The first vector of the tetrad is tangent to such a curve,
\begin{align}
	\frac{dx^\mu}{d\tau}=l^\mu, \qquad \al^2=0,
\end{align}
and the photon acceleration is
\begin{align}
	\aw=\nabla_\al\al,
\end{align}
while the second vector $\an$ satisfies
\begin{align}
	\an^2=0, \qquad \an\cdot\al=-1.
\end{align}
A pair of complex conjugate null vectors,
\begin{align}
	\am \cdot\bar\am=1, \quad \al\cdot \am=0, \quad \an\cdot\am=0,
\end{align}
completes the tetrad.\ It can be constructed from a pair of spacelike vectors $\ave_{1,2}$ as in Eq.~\eqref{eq:m.mbar}.

The relations between tetrad vectors are preserved under three distinct classes of transformations \cite{C:92}, namely transformations that
\begin{enumerate}[label=\Roman*.]
	\item leave the vector $\al$ unchanged, i.e., $\al\to \al$ and
	\begin{equation}
		\am \to \am + a \al, \; \bar\am\to \bar\am+a^*\al, \; \an\to\an+a^*\am+a\bar\am+|a|^2\al . \tag{A.I}
		\label{app:eq:A.I}
	\end{equation}
	\vspace*{-6mm}
	\item leave the vector $\an$ unchanged, i.e., $\an\to \an$ and
	\begin{equation}
		\am \to \am + b \an, \; \bar\am\to \bar\am+b^*\an, \; \al\to\al+b^*\am+b\bar\am+|b|^2\an . \tag{A.II}
		\label{app:eq:A.II}
	\end{equation}
	\vspace*{-6mm}
	\item leave the directions of $\al$ and $\an$ unchanged and rotate the vectors $\ave_{1,2}$ by an angle $\theta$ in the $(\ave_1,\ave_2)$-plane, i.e.,
	\begin{align}
		& \al\to A^{-1}\al, \; \an\to A\an , \tag{A.IIIa} \label{app:eq:A.IIIa} \\
		& \am\to e^{i\theta}\am, \; \bar\am\to e^{-i\theta}\bar\am . \tag{A.IIIb} \label{app:eq:A.IIIb}
	\end{align}
\end{enumerate}
Since $\al$ is defined as the tangent to a specific null curve, transformations of class II are inapplicable in our case. In Ref.~\cite{F:20}, the transformations \eqref{app:eq:A.I}, \eqref{app:eq:A.IIIa}, and \eqref{app:eq:A.IIIb} are referred to as (2), (1), and (3), respectively.

Given an observer with a four-velocity $\au$, $\au^2=-1$, the observed frequency is $\omega_\text{obs}=-u_\mu k^\mu=-\omega u_\mu l^\mu$ [cf.\ Eq.~\eqref{wav}]. The vector $\an$ may be then fixed by
\begin{align}
	\an=-\frac{1}{\au\cdot\al}\left(\au+\frac{\al}{2\au\cdot\al}\right) ,
\end{align}
relating the choice of the second null vector to the reference frame of the observer \cite{O:20}.

Following Ref.~\cite{F:20}, we use a static observer. The Newton gauge \cite{BT:11,BDT:11} complies with this choice. Specifically, we use it to set the initial three-dimensional directions $\bfe_{1,2}$, and thus the initial values of $\am$, $\bar\am$, and $\an$. For a tetrad propagated according to Eq.~\eqref{tetrad-prop}, the only allowed tetrad transformations are the classes I and III that can be performed at the initial moment \cite{F:20}. The identification of $\am$ and $\bar\am$ with right and left circular polarization vectors, respectively (see App.~\ref{app:shortwave}), excludes transformations of class I. Transformations of type \eqref{app:eq:A.IIIb} add a constant phase, leave the main propagation equation \eqref{eq:master} invariant, and trivially modify Eq.~\eqref{tetrad-prop}. On the other hand, transformations of type \eqref{app:eq:A.IIIa} redefine the proper time.

Labeling one-forms in the same way as the corresponding vectors, the volume form is given by
\begin{align}
	\text{vol}_4 = -i\al\wedge\an\wedge\am\wedge\bar\am=-i\al\wedge\an\wedge\ave_1\wedge\ave_2 .
\end{align}
Thus
\begin{align}
\sqrt{-g}=\epsilon^{\mu\nu\rho\sigma}l_\mu n_\nu {\ave_{(1)\rho}\ave_{(2)\sigma}} ,
\end{align}
where $-\epsilon^{\mu\nu\rho\sigma}=\epsilon_{\mu\nu\rho\sigma}=1$ denotes the Levi-Civita symbol.

\section{Details of the shortwave asymptotics} \label{app:shortwave}
For a vector potential
\begin{align}
	A^{\mu}(x) =\eA^\mu(x)e^{i\Phi(x)},
\end{align}
the field $F=dA$ satisfies
\begin{align}
	&F_{\mu\nu}=i\omega\eF_{\mu\nu}e^{i\Phi}, \quad \eF_{\mu\nu}=\eB_{\mu\nu}-\frac{i}{\omega}\eC_{\mu\nu},\\
	&\eB_{\mu\nu}=l_\mu \eA_\nu-l_\nu \eA_\mu, \quad \eC_{\mu\nu}=\eA_{\nu;\mu}-\eA_{\mu;\nu},
\end{align}
where we adopted the conventions of Ref.~\cite{F:20}. Self-dual and anti-self-dual fields correspond to right- and left-handed circular polarization, respectively. Self-dual fields satisfy the conditions
\begin{align}
	&\eF_{\mu\nu}m^\mu n^\nu=0, \label{pol1}\\
	&\eF_{\mu\nu}(-l^\mu n^\nu-\bar m^\mu m^\nu)=0,\label{pol2} \\
	&\eF_{\mu\nu}l^\mu\bar m^\nu=0,\label{pol3}
\end{align}
and the conditions for anti-self-dual fields are analogous with the replacement $\am \leftrightarrow \bar{\am}$ \cite{F:20}.

The Lorenz gauge condition takes the form
\begin{align}
	i\omega l^\mu\eA_\mu +\eA_\mu^{~;\mu}=0. \label{lg}
\end{align}
All effects that were mentioned above --- light deflection, polarization rotation, and birefringence --- are obtained from the two leading terms of the propagation equation $F_{\mu\nu}^{~~~;\nu}=0$,
\begin{align}
	\omega^2 \eB_{\mu\nu}l^\nu-i\omega\left(\eB_{\mu\nu}^{~~~;\nu}+\eC_{\mu\nu}l^\nu\right)=\cO(\omega^0).
	\label{app:eq:prop2}
\end{align}
Using the gauge condition \eqref{lg}, this equation takes the form
\begin{align}
	\omega^2l_\mu l^\mu \eA_\nu -i\omega\left(2l^\mu\eA_{\nu;\mu}+l^\mu_{~;\mu}\eA_\nu\right)=\cO(\omega^0).
	\label{app:eq:propmtw}
\end{align}
The usual treatment \cite{MTW,BW:99} that results in the propagation of light rays along geodesics and polarization rotation considers $\omega \al=\nabla\Phi$ without expanding it order-by-order in $\omega^{-1}$. Expanding both $l_\mu$ and $\eA_\mu$ brings terms of order $\omega^{-1}$ into the first two terms of the propagation equation and establishes gravity-induced birefringence.

Based on the first two terms in the expansion,
\begin{align}
	l^\mu &=\accentset{\circ}l^\mu+(\omega)^{-1}l_1^\mu+\cO(\omega^{-2}),
	\label{app:eq:lmu} \\
	\eA^\mu &=A_0^\mu+(\omega)^{-1}A_1^\mu+\cO(\omega^{-2}),
	\label{app:eq:Amu}
\end{align}
the Lorenz gauge condition~\eqref{lg} yields two relations, namely
\begin{align}
	\accentset{\circ}l_\mu A_0^\mu &=0,
	\label{propO2} \\
	iA_{0\mu}^{~~;\mu} &=\accentset{\circ}l_\mu A_1^\mu+l_1^\mu A_{0\mu}.
	\label{propO1}
\end{align}
Using these relations in Eq.~\eqref{app:eq:propmtw} and noting that
\begin{align}
	l_{\mu;\nu} = \frac{\Phi_{;\mu\nu}}{\omega} = \frac{\Phi_{;\nu\mu}}{\omega} = l_{\nu;\mu} ,
\end{align}
Eq.~\eqref{app:eq:propmtw} becomes
\begin{align}
	\begin{aligned}
		& \omega^2 \accentset{\circ}\al^2A_{0\mu}-i\omega\left(2\accentset{\circ}l^\nu A_{0\mu;\nu}+\accentset{\circ}l_\nu^{~;\nu}A_{0\mu}\right) \\
		& \quad + \omega\Big(\accentset{\circ}\al^2A_{1\mu}+2 \accentset{\circ}l_\mu l_1^\nu A_{0\nu}\Big)=\cO(\omega^0).
	\end{aligned}
\end{align}
It results in the two conditions
\begin{align}
	& \; \accentset{\circ}\al^2=0, \\
	& 2\accentset{\circ}l^\nu A_{0\mu;\nu}+\accentset{\circ}l_\nu^{~;\nu}A_{0\mu}+ 2i \accentset{\circ}l_\mu l_1^\nu A_{0\nu}=0 , 
\end{align}
which coincide with the standard equations of geometric optics \cite{MTW,F:20,O:20} for $\al_1 \to 0$, $\accentset{\circ}\al \to \al$. Introducing the polarization vector and its magnitude,
\begin{align}
	\aA_0= \mathfrak{a}_0\ave, \qquad \mathfrak{a}_0^2=\aA_0\cdot \aA_0,
\end{align}
leads to Eq.~\eqref{paral} and the propagation equation for $\mfa_0$. The polarization equations \eqref{pol1}--\eqref{pol3} identify the two complex tetrad vectors with right and left circular polarizations, respectively:
\begin{align}
	\aA_0^+=\mfa_0\am,\qquad \aA_0^-=\mfa_0\bar\am.
\end{align}
The null condition $\al^2=0$ is interpreted as the Hamilton-Jacobi equation, and the Hamiltonian equations of motion for massless null particles are obtained with the Hamiltonian \cite{F:20,O:20}
\begin{align}
	H=\frac{1}{\omega} g^{\mu\nu}p_\mu p_\nu.
\end{align}
The above $\al$ and $\aA$, identified as the leading terms $\accentset{\circ}\al$ and $\aA_0$, respectively,  are compatible with the full set of Eqs.~\eqref{propO2}--\eqref{propO1}. For a right-handed polarization, for instance, it is achieved by taking
\begin{align}
	\al=\accentset{\circ}\al+(\omega^{-1})(\ab_0+\ab_1),
\end{align}
where \cite{F:20}
\begin{align}
	b_0^\mu=i\bar m^\nu m_{\nu}^{~;\mu}.
\end{align}
Using the polarization conditions \eqref{pol1}--\eqref{pol3}, we find that
\begin{align}
	\al\cdot\ab_1=\am\cdot\ab_1=\bar\am\cdot\ab_1=0,
\end{align}
and thus the expression for $\al$ is obtained up to at most a redefinition of the affine parameter \cite{F:20},
\begin{align}
	l_\mu= \accentset{\circ}l_\mu\pm i(\omega^{-1})\bar m^\nu m_{\nu;\mu}.
\end{align}
In particular, $\al^2 \sim \cO(\omega^{-2})$ or higher \cite{O:20,HO:22}. Proceeding analogously with the case of geometric optics, the Hamiltonian equations of motions are derived, leading to Eq.~\eqref{eq:master}

\section{\\ Properties of null geodesics in the Schwarzschild spacetime} \label{schw}
In what follows, we adapt the conventions and expressions of Ref.~\cite{C:92}. Using the conservation law, setting $\varepsilon=1$, adapting the convention that the geodesic motion occurs in the equatorial plane, and making the polar angle a decreasing function of the affine parameter with the perihelion at $\phi=0$, the Lagrangian ${\cal{L}}=-\tfrac{1}{2} g_{\mu\nu}\dot x^\mu\dot x^\nu$ leads to the equations of motion that include, in particular
\begin{align}
	\dot{\rho}^2 = 1 - f\frac{b^2}{\rho^2}, \qquad \dot{\phi} = - \frac{b}{r_\sg\rho^2} .
\end{align}
With the help of the auxiliary parameters
\begin{align}
	q &\defeq \sqrt{(b_1-1)(b_1+3)}, \\
	k & \defeq \sqrt{\frac{1}{2q}(q-b_1+3)} ,
\end{align}
the three roots of $l^r=p$ are $b_1$ and
\begin{align}
	b_{2,3} = \frac{2b_1}{b_1-1\mp q} .
\end{align}
For $b>b_\text{cr}$, the three roots are real and $b_2<0$.

The trajectory is parametrized by
\begin{align}
	\rho &= \left(\frac{1}{b_1}-\frac{q-b_1+3}{4b_1}(1+\cos\chi)\right)^{-1}, \\
	\phi &= 2 \sqrt{\frac{b_1}{q}} \left[ K(k)-F\big(\half\chi,k\big) \right],
\end{align}
where $\chi=\pi$ corresponds to the perihelion, and $F$ and $K$ denote the Jacobi elliptic integral of the first kind and the complete elliptic integral of the first kind, respectively. The limit $\rho\to\infty$ corresponds to $\chi\to\chi_\infty$, where
\begin{align}
	\sin^2\half \chi_\infty=\frac{q-p+1}{q-p+3}.
	\label{chinf}
\end{align}

\section{\\ Procedure for obtaining iterative analytical solutions} \label{sis}
Here, we illustrate the general discussion of Sec.~\ref{solutions} by explicitly constructing $\vartheta_0$ --- the first term in the iterative solution.
\begin{widetext}
	The evaluation of Eq.~\eqref{q-in} for $j=0$ results in
	\begin{align}
		\begin{aligned}
			q^\rin_0 &= \frac{\sigma}{\omega r_\sg} \left[ \frac{b}{4\rho^2}  \left(\frac{2\mathrm{arccsc} b_1}{\sqrt{b_1^2-1}} - \frac{1}{b_1^2}\right) + \frac{b}{24\rho^2} \left(\frac{4b^2}{b_1^2\rho^3} + \frac{15}{\rho^2} - \frac{6}{\rho} + \frac{6(b^2-b_1^2\rho)\sqrt{\rho^2-b_1^2}}{b_1^4\rho^2} \right. \right. \\
			& \quad \left. \left. \; + \; 6 \ln \frac{\rho}{\rho-1} + \frac{12(b_1^4-b^2)}{b_1^5} \arctan \frac{\rho-\sqrt{\rho^2-b_1^2}}{b_1} + \frac{12}{\sqrt{b_1^2-1}} \arctan \frac{1-\rho+\sqrt{\rho^2-b_1^2}}{\sqrt{b_1^2-1}} \right) \right] .
		\end{aligned}
	\end{align}
 \end{widetext}
This expression already contains the higher powers of $1/b$ that are justified if the approximation $p(\rho)\approx p_0(\rho)$ is used. These terms were retained in the intermediate calculations, but not in the final results.  Thus,
\begin{equation}
	\vartheta_0^\rin = \frac{3\sigma}{8 \omega r_\sg b_1^2 \rho^2} \left( \sqrt{\rho^2/b_1^2-1} - \rho^2\arctan\sqrt{\rho^2/b_1^2-1} \right) .
\end{equation}
The remaining expressions are obtained analogously and are given explicitly in the \textsc{GitHub} repository listed as Ref.~\cite{GitHubRep}. Table~\ref{tab:comparison.analytical.numerical} compares our iteratively obtained analytical solutions to our numerical solutions.

\begin{table}[!htpb]
	\begin{minipage}{\linewidth}
	\centering
	\begin{tabular*}{\linewidth}{@{\extracolsep{\fill}}c cc cc}
	\toprule \toprule
	\small{$b_1$} & \small{$\vartheta (b_1) [\text{rad}]$} & \small{$S_7(b_1) [\text{rad}]$} & \small{$\vartheta^\rou(\infty) [\text{rad}]$} & \small{$S^\rou_7(\infty) [\text{rad}]$} \\
	\midrule \\[-3mm]
  	\small{10} & \small{$-5.26 \times 10^{-3}$} & \small{$-5.25 \times 10^{-3}$} & \small{$2.15 \times 10^{-3}$} & \small{$2.02 \times 10^{-3}$}  \\[1ex]
 	\small{50} & \small{$-2.01 \times 10^{-4}$} & \small{$-2.01 \times 10^{-4}$} & \small{$1.54 \times 10^{-5}$} & \small{$1.46 \times 10^{-5}$}  \\[1ex]
 	\small{$10^{2}$} & \small{$-5.02 \times 10^{-5}$} & \small{$-5.01 \times 10^{-5}$} & \small{$1.91 \times 10^{-6}$} & \small{$1.85 \times 10^{-6}$}  \\[1ex]
	\small{$10^{3}$} & \small{$-5.00 \times 10^{-7}$} & \small{$-5.00 \times 10^{-7}$} & \small{$1.88 \times 10^{-9}$} & \small{$1.82 \times 10^{-9}$}  \\[1ex]
	\small{$10^{5}$} & \small{$-5.00 \times 10^{-11}$} & \small{$-5.00 \times 10^{-11}$} & \small{$1.88 \times 10^{-15}$} & \small{$1.79 \times 10^{-15}$}  \\[1ex]
	\bottomrule \bottomrule
	\end{tabular*}
  	\caption{Comparison of iteratively obtained analytical expressions $S_7=\sum_{j=0}^7 \vartheta_j$ and numerical solutions $\vartheta$. All quoted $\vartheta$ values are calculated for the parameter choice $r_\sg = \sigma = \omega = 1$. Discrepancies in the numerical solutions can be attributed to the fact that the evaluation of the outgoing part of the trajectory requires cancelations or near-cancelations of rather large quantities, thus modifying the scaling behavior from $1/b^2$ to $1/b^3$ and changing the sign.}
  	\label{tab:comparison.analytical.numerical}
  	\end{minipage}
  	\vspace*{-3mm}
\end{table}

We now demonstrate that the $N \to \infty$ limit of the partial sums $S_N$ [cf.\ Eq.~\eqref{eq:vartheta.partialsum.decomposition}] is finite. Due to the initial conditions that set the initial deviation and its rate of change to zero, the inhomogeneous term $w_j=-b^2/(r_\sg \rho^4)\vartheta_{j-1}$ and $w_{j+1}$ have opposite signs for $j \geqslant 1$ due to their dependence on $\vartheta_{j-1}$ and $\vartheta_{j}$, respectively. We show that either $|\vartheta_{j+1}|<|\vartheta_{j}|$ or $|\vartheta_{j+1}|<|\vartheta_{j-1}|$, and thus $S_N$ is convergent at each point by virtue of the Leibniz convergence criterion. For simplicity, we only consider the ingoing segment of the trajectory in what follows. The outgoing segment is treated analogously.

From Eq.~\eqref{q-in}, it follows that
\begin{align}
	\begin{aligned}
 		|q^\rin_j(\rho)|&<|\vartheta_{j-1}^{\text{max}}|\frac{b^2}{r_\sg \rho^2}\left| \int_\rho^\infty\frac{d\rho}{\rho^2p_0(\rho)} \right| \\
 		&\leqslant|\vartheta_{j-1}^{\text{max}}|\frac{b^2  }{r_\sg b_1\rho^2}\left(\half{\pi}-\arctan|p_0|\right),
 	\end{aligned}
\end{align}
and therefore
\begin{align}
	|\vartheta^\rin_j(\rho)|<|\vartheta_{j-1}^{\text{max}}|\frac{b^2}{8 b_1^2}\big(\pi-2\arctan|p_0|\big)^2.
	\label{step1}
\end{align}
Continuing the iterations, we have
\begin{align}
  |q^\rin_{j+1}(\rho)|<|\vartheta_{j-1}^{\text{max}}|\frac{b^4  }{48 b_1^3 r_\sg \rho^2}\big(\pi-2\arctan|p_0|\big)^3
\end{align}
and
\begin{align}
	|\vartheta^\rin_{j+1}(\rho)|<|\vartheta_{j-1}^{\text{max}}|\frac{b^4  }{384 b_1^4}\big(\pi-2\arctan|p_0|\big)^4. \label{step2}
\end{align}
Equation~\eqref{step1} establishes that $|\vartheta^\rin_j(\rho)|<|\vartheta_{j-1}^{\text{max}}|$ outside of a certain neighborhood of $b_1$. On the other hand, from Eq.~\eqref{step2} it follows that $|\vartheta^{\text{max}}_{j+1}(\rho)|<|\vartheta_{j-1}^{\text{max}}|$. If $|\vartheta^{\text{max}}_{j}(\rho)|<|\vartheta_{j-1}^{\text{max}}|$, then the convergence is established by assumption. If $|\vartheta^{\text{max}}_{j}(\rho)|>|\vartheta_{j-1}^{\text{max}}|$, then Eq.~\eqref{step2} implies that
\begin{align}
	|\vartheta^{\text{max}}_{j+1}(\rho)|<|\vartheta_{j-1}^{\text{max}}|<|\vartheta_{j}^{\text{max}}|,
\end{align}
again establishing the convergence of $S_N$.\ In App.~\ref{app:full.expr}, we present explicit expressions for order-by-order iterations of $\gb$, $\gc_\infty$, and $\gc$, which show how the series converges in more detail.

\vspace*{7mm}
\section{\\ Full expressions for order-by-order iterations} \label{app:full.expr}
\vspace*{-3.6mm}
The order-by-order iterations of $\gb$, $\gc_\infty$, and $\gc$ are given explicitly by
\begin{align}
	\gb_{(N)} \defeq \sum_{j=0}^N \gb_j ,
\end{align}
etc., for various values of $N$ that are determined by the convergence speed. In particular,
\begin{widetext}
	\vspace*{-4.22mm}
	\begin{align}
		\gb_\text{(4)} = \left(\frac{3\pi }{16}\right)_{(0)} & \! + \left(\frac{3 \pi}{64}-\frac{\pi ^3}{128}\right)_{(1)} \! + \left(\frac{3 \pi }{256}-\frac{\pi ^3}{512}+\frac{\pi ^5}{10240}\right)_{(2)} \! + \left(\frac{3 \pi }{1024}-\frac{\pi ^3}{2048} + \frac{\pi ^5}{40960}-\frac{\pi ^7}{1720320}\right)_{(3)} \nonumber \\
		 & \! + \left(\frac{3 \pi }{4096}-\frac{\pi ^3}{8192}+\frac{\pi ^5}{163840}-\frac{\pi ^7}{6881280}+\frac{\pi ^9}{495452160}\right)_{(4)} \!  .
	\end{align}
	Numerically,
	\begin{align}
		\gb_\text{(4)} &= (0.58905)_{(0)} + (-0.09497)_{(1)} + (0.00614)_{(2)} + (-0.00022)_{(3)} + (0.00001)_{(4)} .
	\end{align}
	Similarly,
	\begin{align}
		\gc_{\infty,\text{(4)}} &= \left(-\frac{1}{3}-\frac{9 \pi }{16}+\frac{3 \pi ^2}{16}\right)_{(0)} \! + \left(-\frac{10}{27}-\frac{15 \pi }{64}+\frac{25 \pi ^2}{192}+\frac{5 \pi ^3}{32}-\frac{\pi ^4}{64}\right)_{(1)} \! + \left(-\frac{91}{243}-\frac{21 \pi }{256}+\frac{721 \pi ^2}{6912}+\frac{7 \pi ^3}{128}-\frac{25 \pi ^4}{2304} \right. \nonumber \\
 		& \left. \quad - \; \frac{7 \pi ^5}{640}+\frac{\pi ^6}{1920}\right)_{(2)} \! + \left( -\frac{820}{2187}-\frac{27 \pi }{1024}+\frac{24025 \pi ^2}{248832}+\frac{9 \pi ^3}{512}-\frac{721 \pi ^4}{82944} - \frac{9 \pi ^5}{2560}+\frac{5 \pi ^6}{13824}+\frac{3 \pi ^7}{8960}-\frac{\pi ^8}{107520}\right)_{(3)} \nonumber \\
 		& \quad + \left( \! -\frac{7381}{19683}-\frac{33 \pi }{4096}+\frac{846241 \pi ^2}{8957952}+\frac{11 \pi ^3}{2048}-\frac{24025 \pi ^4}{2985984}-\frac{11 \pi ^5}{10240}+\frac{721 \pi ^6}{2488320}+\frac{11 \pi ^7}{107520}-\frac{5 \pi ^8}{774144}-\frac{11 \pi ^9}{1935360} \right. \nonumber \\
 		& \left. \quad + \; \frac{\pi ^{10}}{9676800}\right)_{(4)} \! ,
	\end{align}
	where higher-order terms have been omitted as they become increasingly cumbersome, and
	\begin{align}
		\gc_{\infty,\text{(6)}} = (-0.2499)_{(0)} + (3.5011)_{(1)} + (-1.8103)_{(2)} + (0.3883)_{(3)} + (-0.0474)_{(4)} + (0.0038)_{(5)} + (-0.0002)_{(6)} .
	\end{align}
	Lastly,
	\begin{align}
		\gc_\text{(1)} = \left( \frac{3 \pi }{16}\right)_{(0)} + \left( \frac{3 \pi }{64}-\frac{\pi ^3}{128}\right)_{(1)} ,
		\label{app:eq:frakc.analytical}
	\end{align}
	and
	\begin{align}
		\gc_\text{(1)} = (0.58905)_{(0)} + (-0.09497)_{(1)} .
	\end{align}
	Since $\gd$ is obtained when solving for $\rho_0$, where $\rho_0$ corresponds to $\sum_{j=0}^N \vartheta_j(\rho) =0$, we do not expand it order-by-order. Up to the sixth iteration, we find
	\begin{align}
		\gd_{(6)} = \frac{900159609888 \pi ^2}{6434525863463} - \frac{74999564640 \pi ^4}{6434525863463} + \frac{12490769984 \pi ^6}{32172629317315} - \frac{44479072 \pi ^8}{6434525863463} + \frac{195168281 \pi ^{10}}{2573810345385200} .
	\end{align}
\end{widetext}

\end{document}